\documentclass[english,letterpaper,12pt]{article}

\usepackage{fourier}
\usepackage[T1]{fontenc}

\usepackage[margin=1in]{geometry}
\usepackage{amssymb}
\usepackage{amsmath}
\usepackage{amsthm}
\usepackage{mathrsfs}
\usepackage{xfrac}
\usepackage{bbm} % For double-stroke unity
\usepackage{scrextend} % For \addmargin{}
\usepackage{natbib} 
\usepackage{graphicx} % Allows inclusion of figures
\usepackage{color}
\usepackage[table,xcdraw]{xcolor}
\usepackage{multirow} % Allows merged rows in tables
\usepackage{booktabs} % Allows the use of \toprule, \midrule and \bottomrule in tables
\usepackage[labelsep=period]{caption} % Allows `\\' in captions
\usepackage[normalem]{ulem} % Allows crossing out text
\usepackage[hang,flushmargin]{footmisc}
\usepackage{authblk}
\usepackage{scalefnt}
\usepackage[hidelinks]{hyperref}
\usepackage{rotating} % Allows landscape-oriented tables
\usepackage{setspace} % Allows doublespacing
\usepackage[compact]{titlesec} % Allows section title formatting

\numberwithin{equation}{section}

\allowdisplaybreaks

\DeclareMathOperator*{\argmin}{argmin}
\DeclareMathOperator*{\argmax}{argmax}

\makeatletter
\renewcommand{\paragraph}{%
	\@startsection{paragraph}{4}%
	{\z@}{1.2ex \@plus 1ex \@minus .2ex}{-1em}%
	{\normalfont\normalsize\bfseries}%
}
\makeatother

\newtheoremstyle{style}
	{7pt} % Space above
	{5pt} % Space below
	{} % Body font
	{} % Indent amount
	{\bfseries} % Theorem head font
	{.} % Punctuation after theorem head
	{.5em} % Space after theorem head
	{} % Theorem head spec (can be left empty, meaning `normal')
\theoremstyle{style}
\newtheorem{remark}{Remark}

\begin{document}
	
\title{\huge On the Estimation of Cross-Firm Productivity \\ Spillovers with an Application to FDI\thanks{\textit{Email}: emir.malikov@unlv.edu (Malikov) and shunanzhao@oakland.edu (Zhao). \newline\newline We benefited greatly from feedback and discussions with Paul Grieco, Jordi Jaumandreu, Devesh Raval, Chris Vickers, Nic Ziebarth, seminar participants at Auburn, Binghamton, Nebraska, TAMU and UNLV as well as  participants at the 2019 Midwest Econometrics Group meeting at the Ohio State and the North American Productivity Workshop XI at the University of Miami.} }

\author[1]{\sc Emir Malikov}
\author[2]{\sc Shunan Zhao}
\affil[1]{\small University of Nevada, Las Vegas}
\affil[2]{\small Oakland University}

\date{\small June 20, 2021}
\maketitle

%\vspace{\baselineskip}

\begin{abstract}
\noindent \small We develop a novel methodology for the proxy variable identification of firm productivity in the presence of productivity-modifying learning and spillovers which facilitates a unified "internally consistent" analysis of the spillover effects between firms. Contrary to the popular two-step empirical approach, ours does not postulate contradictory assumptions about firm productivity across the estimation steps. Instead, we explicitly accommodate cross-sectional dependence in productivity induced by spillovers which facilitates identification of both the productivity and spillover effects therein simultaneously. We apply our model to study cross-firm spillovers in China's electric machinery manufacturing, with a particular focus on productivity effects of inbound FDI.

\vspace{\baselineskip}



\end{abstract}

\doublespacing %\setlength{\baselineskip}{24pt} \renewcommand{\baselinestretch}{1}
\setlength{\abovedisplayskip}{3pt}
\setlength{\belowdisplayskip}{3pt} 
\thispagestyle{empty} \addtocounter{page}{0}
\clearpage

% ------------------------------------------------------------------------------------------
% ------------------------------------------------------------------------------------------

\section{Introduction}\label{sec:intro}

Since its popularization by \citet{marshall1890}, the concept of cross-firm knowledge, or technology, spillovers has increasingly become a central fixture in many economic theories, including of long-run growth, spatial agglomeration, research and innovation, international trade and more. The idea is simple: firms improve their productivity by learning from one another, with the most commonly conjectured drivers of these knowledge exchanges (technology transfers) being human interaction along with spatial and industrial/technological proximity. These productivity spillovers can also propel significant positive externalities in many productivity-enhancing activities such as research and development (R\&D), foreign direct investment (FDI) or exporting. In this paper, we develop a new methodology for the proxy variable structural identification of firm productivity in the presence of productivity-modifying learning and spillovers which facilitates a unified ``internally consistent'' analysis of the spillover effects between peer firms.

Although productivity is straightforward in concept, its measurement is not trivial for a multitude of reasons  among which is the inherent latency of firm productivity/efficiency. Naturally, the identification of productivity spillovers across firms is even more challenging a task because, as \citet[][p.53]{krugman1991} points out, ``knowledge flows ...~are invisible; they leave no paper trail by which they may be measured and tracked.'' On this account, most empirical work on cross-firm technology spillovers abstracts away from pinpointing specific mechanisms by which such spillovers occur\footnote{Nuanced empirical studies of spillovers are rare and require detailed matched (and usually proprietary or confidential) datasets and, by design, have a limited identifying ability restricted to particular channels/mechanisms of knowledge diffusion such as labor turnover \citep[e.g.,][]{balsvik2011,poole2013,stoyanovzubanov2012} or coauthorship networks \citep[e.g.,][]{zacchia2020}. Others resort to limiting the scope of analyzed spillovers: e.g., \citet{jiangetal2018} restrict FDI spillovers to direct links between joint ventures and their domestic partners, whereas \citet{newman2015} focus on  productivity spillovers along vertical supply chains.} and instead focuses on a simpler but more feasible objective of testing for the presence of cross-firm spillovers in general. The most common frameworks either (i) focus squarely on ``productivity spillovers'' by seeking to identify how a firm's productivity is affected by that of its peers\textemdash the ``endogenous effect'' in the \citet{manski1993} nomenclature\textemdash or (ii) take a more reduced-form approach centered only on measuring ``contextual'' spillover effects of various productivity-modifying activities (FDI, R\&D, exporting, etc.) facilitated by cross-firm spillovers in productivity. Recent examples of the first include \citet{bazzietal2017} and \citet{serpakrishnan2018} who study vertical productivity spillovers along supply chains and material-product connections. As it happens, the literature embracing the second framework is more predominant and has a longer history: e.g., see \citet{alvarez2008} on spillovers from exporting; \citet{javorcik2004}, \citet{javorcikspatareanu2008}, \citet{haskeletal2007}, \citet{blalockgertler2008},  \citet{kelleryeaple2009}, \citet{barriosetal2011}, \citet{luetal2017} on FDI spillovers;  \citet{branstetter2001}, \citet{griffithetal2006},  \citet{bloometal2013},  \citet{zacchia2020} on technology spillovers from R\&D; and \citet{acharyakeller2008} on productivity spillover effects of imports.

Both empirical frameworks are usually operationalized in two steps, whereby one first recovers firm productivity from the production function estimates and then examines spillovers in the second step by (linearly) regressing these productivity estimates on various peer-group averages capturing firms' exposure to potential spillovers. Owing to its popularity and ease of implementation, most studies estimate firm productivity in the first step via the proxy variable approach \`a la \citet{op1996} and \citet{lp2003} that typically assumes that each firm's productivity process is an independent (over firms) exogenous Markov chain. However, if present, spillovers would generate cross-sectional dependence among firms, which is nonetheless being overlooked in the first-step estimation of productivity. Not only does this raise reservations about the identification of production function (and hence, productivity) econometrically, but more importantly, such a two-step procedure suffers from the conceptual ``internal inconsistency'' because the second-step regressions, in effect, contradictorily postulate the existence of spillover-induced cross-firm dependence in productivity which is at odds with the identifying assumptions used in the first step. As such, conclusions about spillovers based on a two-step procedure may be spurious.

With the above in mind, we provide a novel (semiparametric) methodology for the estimation of productivity spillovers. In line with the existence of cross-firm spillovers, in building our model, we explicitly accommodate cross-sectional peer dependence in firm-specific (latent) productivity that such spillovers induce. This is fundamentally different from the aforementioned traditional two-step approach.\footnote{We should note that a two-step framework is not universal across empirical studies of productivity spillovers. The exceptions are predominantly from the literature on R\&D-borne spillovers that favors the estimation of ``augmented production functions.'' We discuss the benefits of our methodology over the latter in Appendix \ref{sec:appendix_augment}.} 

To keep our methodology amenable to a wide range of contexts, we conceptualize peer dependence in firm performance via spatiotemporal spillovers in latent productivity itself. We generalize the conventional setup of firm production assumed in the literature to introduce the dependence of each firm's productivity on its (geographically and industrially proximate) peer-group average productivity. To that end, we  dispense with the standard assumption of independent (over $i$) exogenous Markov process for latent productivity \citep[e.g.,][and others]{op1996,lp2003,acf2015,gnr2013} in favor of a controlled productivity process with explicitly incorporated cross-sectional dependence. This permits the firm to improve its productivity by learning not only directly from its own productivity-modifying activities but also indirectly from the activities of its peers.

Explicit modeling of cross-sectional dependence in firm productivity directly affecting its evolution (along with a structural timing assumption about learning process) enables us to build upon the popular proxy variable technique to develop a unified identification scheme for both the latent productivity and spillover effects therein \textit{simultaneously} that is also robust to \citeauthor{acf2015}'s (2015) and \citeauthor{gnr2013}'s (2020) critiques. In fact, as we show in the paper, estimating the firm production function or productivity using traditional proxy methods while ignoring the spillover-induced cross-sectional dependence, as customarily done in the literature, likely leads to misspecification and  omitted variable bias. This underscores the key practical advantage of our proposed methodology. Also, by virtue of a nonparametric formulation of the firm productivity process, we transcend restrictive additively linear specifications favored in the spillovers literature which lets us accommodate heterogeneous spillover effects. %due to the varying exposure to peers, different within-firm learning experiences and the varying absorptive capacity to learn in the first place.

Because our methodology can be easily adapted to admit various spillover origins, it is fit to investigate productivity spillovers in many contexts, including spatial agglomeration, R\&D externalities, learning from exporters, and others. In our paper, for example, we consider an application to the FDI inflows. 

We also contribute to the literature on proxy-based identification of production functions more broadly, by providing a practical, easy-to-implement semiparametric adaptation of \citeauthor{gnr2013}'s (2020) estimator. Our point of departure is a parametric assumption about the functional form of production function, which is the predominant modeling strategy in productivity literature with the Cobb-Douglas specification being the most popular among researchers. Along the lines of \citet{dj2013}, our modeling approach fully embraces the assumed parametric specification of the production function by explicitly utilizing a known functional form of the static first-order condition for materials and the inverse conditional input demand function that it implies. By doing so, we circumvent the need to integrate the estimated material elasticity function at \textit{each} observation in order to recover the unknown production function required by \citeauthor{gnr2013}'s (2020) more computationally demanding, albeit admittedly less restrictive, nonparametric methodology. In contrast, our parametric inversion of the material demand yields a much simpler semiparametric estimator. We also show how to extend our methodology to more flexible specifications of the firm's production function such as translog.

Besides the empirical application of our methodology, we also demonstrate its ability to successfully identify firm productivity and cross-firm spillovers therein in a set of Monte Carlo experiments. The results are encouraging and show that our approach recovers the true parameters well, thereby lending strong support to the validity of our identification strategy. We also use the simulations to show how estimating spillovers via the popular but internally inconsistent two-step procedure can lead to spurious and misleading results.

The rest of the paper unfolds as follows. Section \ref{sec:fdi} provides context for our  application centered on FDI spillovers. Section \ref{sec:model} describes a generic model of firm production with productivity-modifying learning and spillovers. We discuss identification and estimation in Sections \ref{sec:identification} and \ref{sec:estimation}, respectively. Section \ref{sec:simulations} reports simulation results. We present our empirical application in Section \ref{sec:fdiapplication} and conclude in Section \ref{sec:conclusion}.

% ------------------------------------------------------------------------------------------
% ------------------------------------------------------------------------------------------

\section{Application to FDI}
\label{sec:fdi}

Cross-firm productivity spillovers can propel significant positive externalities in many productivity-enhancing activities, which are especially important from a policy perspective. Take, for instance, inbound foreign direct investment. 

Public policies aimed at attracting FDI are commonplace both in developing and developed economies. Besides immediate returns in the form of capital inflows and employment gains, the primary justification for the FDI-promoting government incentives is mostly centered around gaining access to intangible productive ``knowledge'' assets from abroad such as new technologies, proprietary know-hows, more efficient and innovative marketing and management practices, established relational networks, reputation, etc., which can boost productivity of domestic firms. More crucially, productivity-enhancing effects of inbound FDI are widely believed to realize broadly beyond immediate recipients, who benefit from direct learning of foreign knowledge, by also benefiting many other domestic firms via productivity spillovers. These spillovers may occur via informal contacts (e.g., attendance of trade shows, exposure to affiliate and/or competitor products and marketing, learning by imitation, customer-supplier discussions), more formal reverse engineering, or labor turnover, eventually yielding large within- and/or cross-industry productivity gains. Measuring the extent and significance of these ``social returns'' of FDI is therefore imperative for the design of effective industrial policy. 

To empirically showcase our estimator, we apply it to study horizontal productivity spillovers in China's electric machinery manufacturing industry in 1998--2007, with a particular focus on the technology-transfer effects of inbound FDI on productivity via domestic firms' learning of more advanced/efficient foreign knowledge to which they may gain access directly through their \textit{own} foreign investors and indirectly through spillovers from their foreign-invested \textit{peers}. Among the world's top destinations for foreign investment, China presents a natural environment for the analysis of broad productivity effects of FDI on domestic firms  especially because of its ``open door'' policies aimed at promoting foreign investment (e.g., special economic zones with regulatory environments favorable to foreign capital) and its fairly recent accession to the World Trade Organization in 2001. Focus on the electric machinery industry in particular is motivated by its being historically one of China's most fundamental manufacturing sectors and among the largest FDI recipients (see Appendix \ref{sec:indsutry} for more on the choice of the industry).

The empirical literature on FDI spillovers has generally produced mixed findings, especially for the long-sought-after horizontal productivity spillovers \citep[see][for excellent surveys]{keller2008,keller2010}. 
Few earlier studies that have analyzed external productivity spillovers from inbound FDI in China \citep[see][and the references therein]{ jiangetal2018} have done so using the two-step approach or by ``augmenting'' the firm's production function with the associated methodological issues as discussed earlier. The results have been mixed, further heightening the appeal of our study based on a new methodology. The reanalysis of FDI-borne technology spillovers in China is also timely and relevant in light of the ongoing trade disputes between the U.S.~and China fostered, among other things, by grievances of the former against China's ``unfair technology transfer regime'' for foreign companies. Investigating the extent of external spillovers can therefore provide an informative context for a more holistic understanding of the FDI environment in the country and the implications of its technology-transfer rules and regulations. 

To briefly preview our key results, we find that at least 87\% of manufacturers of electric machinery in China enjoy significant productivity-boosting effects of inbound FDI, both directly and indirectly. At the median, an increase of the foreign share in all firms' equity by 10 percentage points, in the short run, improves each firm's productivity by 1.4\% via direct learning and by 0.4\% via external effects. The latter indirect effect of FDI is facilitated by substantial cross-firm productivity spillovers in the industry, with the median spillover elasticity estimated at 0.33. These productivity spillovers are significantly positive for about 84\% of firms in the industry.

% ------------------------------------------------------------------------------------------
% ------------------------------------------------------------------------------------------

\section{Production with Learning and Spillovers}
\label{sec:model}

We now describe a generic paradigm of production in the presence of productivity-modifying learning and spillovers. Consider the production process of a firm $i=1,\dots,n$ in time period $t=1,\dots,T$ in which physical capital $K_{it}$, labor $L_{it}$ and an intermediate input such as materials $M_{it}$ are transformed into the output $Y_{it}$ via production function, given the log-additive Hicks-neutral firm productivity. Following the popular convention in the literature \citep[e.g.,][]{op1996,lp2003,topalovakhandelwal2011,dj2013}, we assume that the firm's stochastic production process is Cobb-Douglas:
\begin{equation}\label{eq:prodfn}
	Y_{it} = K_{it}^{\beta_K}L_{it}^{\beta_L}M_{it}^{\beta_M} \exp\left\{\omega_{it}+\eta_{it}\right\} ,
\end{equation}
where the exponent $\omega_{it}+\eta_{it}$ is the latent ``composite'' productivity residual consisting of (i) the firm $i$'s persistent productivity $\omega_{it}$ and (ii) a random transitory productivity shock $\eta_{it}$. Our methodology can also adopt more flexible specifications of the firm's production function such as the log-quadratic translog specification. See Appendix \ref{sec:appendix_translog} for this extension of our model.

We assume that $K_{it}$ and $L_{it}$ are subject to adjustment frictions (e.g., time-to-install, hiring and training costs) and thus are quasi-fixed, whereas $M_{it}$ is freely varying. That is, $M_{it}$ is chosen in period $t$, whereas $K_{it}$ and $L_{it}$ are determined in period $t-1$. Both $K_{it}$ and $L_{it}$ are the state variables with dynamic implications and follow their respective deterministic laws of motion: 
\begin{equation}\label{eq:k_lawofmotion}
K_{it}=I_{it-1}+(1-\delta)K_{it-1}\quad \text{and}\quad L_{it}=H_{it-1}+L_{it-1} ,
\end{equation}
where $I_{it}$, $H_{it}$ and $\delta$ are the gross investment, net hiring and the depreciation rate, respectively. The firm maximizes a discounted stream of expected life-time profits in perfectly competitive output and factor markets. Also, for convenience, let $\mathcal{I}_{it}$ denote the information set available to the firm $i$ for making period $t$ production decisions. 

Our main objective is to study the role of learning and spillovers in the evolution of firm productivity. To that end, we need to dispense with the standard assumption of \textit{exogenous} Markov process for $\omega_{it}$ in favor of a \textit{controlled} productivity process and, more importantly, to explicitly recognize the potential for cross-sectional dependence therein.
For generality sake, we denote the productivity-modifying ``controls'' via a (vector of) generic variable(s) $G_{it}$. This variable may measure the firm's deliberate activities aimed at improving its productivity such as R\&D expenditures \citep{dj2013} or some other aspects of its behavior in the marketplace that have productivity implications such as exporting \citep{deloecker2013}. Depending on the application of interest, $G_{it}$ may also admit measures of the firm's exposure to technological innovations from investors or partners\textemdash the focus of our empirical illustration\textemdash or its access to public subsidies and other forms of favorable treatment from the government owing to political connections, etc. In the end, no matter the choice of $G_{it}$, the rationale for its inclusion in the firm productivity evolution is to capture within-firm ``learning'' facilitated by the firm's own productivity-modifying activities or characteristics. 

Next, we permit the firm $i$ to improve its productivity by learning not only from its \textit{own} activities but also from its \textit{peer} firms. We do so by relaxing the usual assumption of firm productivity being an independent (over $i$) Markov chain to allow for cross-sectional dependence. More concretely, we assume that the $i$th firm's productivity $\omega_{it}$ evolves according to the following controlled first-order process:
\begin{equation}\label{eq:proddist_a_exp}
\omega_{it} = \mathbb{E}\Bigg[\omega_{it} \Bigg|\ \omega_{i,t-1}, G_{i,t-1}, \sum_{j(\ne i)}s_{ij,t-1}\omega_{j,t-1}\Bigg] + \zeta_{it},
\end{equation}
where $\{s_{ij,t-1};\ j(\ne i)=1,\dots,n\}$ are the peer-identifying weights (from the perspective of firm $i$), and $\zeta_{it}$ is a mean-independent unanticipated random innovation in persistent productivity normalized to have a zero mean: $\mathbb{E}[\zeta_{it} | \mathcal{I}_{i,t-1}] = \mathbb{E}[ \zeta_{it} | \omega_{i,t-1}, G_{i,t-1}, \sum_{j(\ne i)}s_{ij,t-1}\omega_{j,t-1}] = 0$.

While the exact choice of how to construct peer weights $\{s_{ijt}\}$ depends on the empirical context, for a general baseline case here, we let the peers be identified based on their spatial vicinity and industrial similarity to firm $i$. Thus, letting $\mathcal{L}(i,t)$ represent a set of spatially proximate ``neighbors'' of the firm $i$ in time period $t$ that also operate in the same industry, peer weights $\{s_{ijt}\}$ are constructed for each $(i,t)$ as follows:
\begin{equation}\label{eq:weight1}
s_{ijt}= \frac{\mathbbm{1}\big\{(j,t)\in \mathcal{L}(i,t)\big\} }
{\sum_{k(\ne i)=1}^{n} \mathbbm{1}\big\{(k,t)\in \mathcal{L}(i,t)\big\} },
\end{equation}	
where the normalization in the denominator yields a convenient interpretation of $\sum_{j(\ne i)}s_{ijt}\omega_{jt}$ as the average peer productivity. Focusing on geographically proximate peers within the industry fits a broader narrative in regional and urban economics about the scopes of agglomeration economies and the localized cross-firm productivity spillovers (due to technology and knowledge diffusion, labor market interactions, etc.) being one of the main sources of such externalities \citep[e.g., see][]{duranton2004,rosenthal2004}. By restricting the scope to the same industry we effectively focus on intra-industry horizontal productivity spillovers.

\begin{remark} \normalfont\label{rem:s_sym} The weighing scheme in \eqref{eq:weight1} treats cross-firm spillovers symmetrically in that all members of a peer group affect each other's productivity. Not only is this a standard approach to measuring ``peer effects,'' but in doing so we also wish to remain as agnostic about spillovers as possible and avoid imposing priors about the directionality of peer dependence. But should one choose to regulate the direction of spillovers by restricting them to occur, say, from more productive to less productive firms only, our framework can be modified to accommodate that too. For more discussion, see Appendix \ref{sec:appendix_asym}.
\end{remark}

\begin{remark} \normalfont\label{rem:s_geo_def} In \eqref{eq:weight1}, a uniform weighting is applied across all peers of the firm $i$ that are located in its spatial proximity. This implicitly assumes that within boundaries of the firm's spatial ``neighborhood''  the distance gradient is of second-order importance for knowledge spillovers. The main benefit of postulating such a feature of peer networks is that it does not require granular geographic data about individual firms and can be operationalized using coarse location information such as ZIP code, census track, city, region. The degree to which this is a reasonable weighting scheme obviously depends on the selected ``level'' of neighborhoods as well as the application-specific institutional context. If desired and feasible, peer weights $\{s_{ijt}\}$ can be appended to incorporate a (decay) function of the distance between $i$ and its peers $\{j\}$.
\end{remark}

The innovativeness of our model in the context of a broader literature on the structural proxy variable estimation of production functions is as follows. The ``controlled'' formulation in \eqref{eq:proddist_a_exp} is more general than the most commonly assumed exogenous Markov process \`a la \citet{op1996} whereby $\omega_{it} = \mathbb{E}\left[\omega_{it} | \omega_{i,t-1}\right] + \zeta_{it}$  because it enables the firm to influence the evolution of its productivity via its own productivity-enhancing activities/characteristics as well as by interacting with other local firms in the industry as captured by $G_{i,t-1}$ and $\sum_{j(\ne i)}s_{ij,t-1}\omega_{j,t-1}$, respectively. While controlled Markov processes for  $\omega_{it}$ are not novel to the literature \citep[e.g.,][]{dj2013,deloecker2013}, all such studies have focused exclusively on an \textit{independently} (over $i$) distributed $\omega_{it}$ having the latter depend on the firm's own productivity and productivity-modifying variables. Our important generalization is that we permit cross-sectional dependence in firm productivity within peer networks due to agglomeration. 

Consider the spatiotemporal autoregressive conditional mean of $\omega_{it}$ in \eqref{eq:proddist_a_exp} that represents the $i$th firm's expected one-period-ahead productivity at time $t-1$. First, by letting it depend on the firm's \textit{own} productivity modifier $G_{i,t-1}$, we are able to account for (internal) \textit{direct} learning taking place within the firm, with the corresponding estimand of interest being
\begin{equation}\label{eq:dl_def}
DL_{it} = \frac{\partial \mathbb{E}[\omega_{it}|\cdot]}{\partial G_{i,t-1}}.
\end{equation}

Second, in including the spatial average of other firms' productivities $\sum_{j(\ne i)}s_{ij,t-1}\omega_{j,t-1}$, not only can we accommodate potential agglomeration externalities facilitated by productivity \textit{spillovers} across firms, but we are also able to capture the (external) \textit{indirect} learning whereby the productivity-modifying activities may have secondary effects on firms beyond their immediate (and intended) beneficiary. Concretely, defining the cross-firm productivity spillovers as
\begin{equation}\label{eq:sp_def}
SP_{it} = \frac{\partial \mathbb{E}[\omega_{it}|\cdot]}{\partial \sum_{j(\ne i)}s_{ij,t-1}\omega_{j,t-1}},
\end{equation}
the measure of firm $i$'s indirect learning from firm $j$'s productivity-modifying activities is 
\begin{align}\label{eq:il_def}
IL_{ijt} &= \frac{\partial \mathbb{E}[\omega_{it}|\cdot]}{\partial G_{j,t-2}} = \frac{\partial \mathbb{E}[\omega_{it}|\cdot]}{\partial \sum_{j(\ne i)}s_{ij,t-1}\omega_{j,t-1}}\left(s_{ij,t-1}\frac{\partial \omega_{j,t-1}}{\partial G_{j,t-2}}\right)= SP_{it} \times s_{ij,t-1}\times DL_{j,t-1}.
\end{align}
The $IL_{ijt}$ effect in \eqref{eq:il_def} is defined for an $(i,j)$ pair of firms, and we can aggregate it to the total indirect learning of firm $i$ from all of its peers as 
\begin{align}\label{eq:til_def}
TIL_{it} &= \sum_{j(\ne i)} IL_{ijt} = SP_{it} \times \sum_{j(\ne i)} s_{ij,t-1} DL_{j,t-1}.
\end{align}

As defined in \eqref{eq:dl_def}--\eqref{eq:til_def}, the learning and spillover effects on firm productivity are ``short-run,'' but they accumulate and diffuse over time owing to a persistent nature of the firm's productivity evolution. Indirectly, this dynamic feature permits two peer firms that are separated temporally to continue to affect one another with the effect size attenuating over time. Such time-separated interactions characterize the temporal scope of productivity spillovers which helps propel \textit{dynamic} agglomeration economies. The underlying idea here is that the knowledge acquired either through internal learning or from peers takes time to accumulate. Together with the geographic and industrial scopes of spillovers embedded in the definition of peer weights $\{s_{ijt}\}$, the autoregressiveness of productivity specification covers the three main dimensions of external economies \citep[see][]{rosenthal2004}.

\begin{remark} \normalfont\label{rem:omegalag_vs_Glag} The total indirect learning effect in \eqref{eq:til_def} is, effectively, a measure of spillovers \textit{specifically} in $G$. In this, our conceptualization of external effects of $G$ as operating through the firm's exposure to the aggregate of its peers' unobservable productivities\textemdash that is, via ``productivity spillovers'' more broadly\textemdash fundamentally differs from the conventional approach to measuring spillovers in productivity-modifying activities (think, FDI, R\&D or export spillovers) that relies on observable industry aggregates of $G$. That is, we measure the $i$th firm's exposure to the \textit{external} knowledge using an aggregate of $\{\omega_{jt};\ j(\ne i)=1,\dots,n\}$ as opposed to an aggregate of $\{G_{jt};\ j(\ne i)=1,\dots,n\}$. Our formulation is more flexible because it does not restrict the origins of cross-firm productivity spillovers to $G$ alone. It is also more realistic and conceptually congruous because it incorporates secondary information about the peer firms' own direct/internal learning facilitated by the productivity-modifying activities they undertake: namely, to learn from one's peers' $G$, peers themselves should learn from their own $G$ first.
\end{remark}

The productivity evolution process in \eqref{eq:proddist_a_exp} characterizes the peer interaction between firms through their productivity. Each firm $i$ has a ``reference group'' of spatially proximate peers from the same industry $\mathcal{L}(i,t)$ with which it interacts. The identification of such cross-peer relations in networks is a notoriously challenging problem \citep[e.g., see][]{manski1993,manski2000,moffitt2001,blumeetal2011}. The potential obstacles include (i) the perfect functional dependence between the average outcome of the group and its mean characteristics due to the so-called ``reflection problem'' which may leave no exogenous variation excluded to instrument the endogenous peer behavior when there is more than one channel for the peer effects, (ii) the confounding presence of unobserved ``correlated'' group effects, and (iii) the endogenous group membership (or network structure). In our case, the additional layer of complexity is the latency of firm productivity. This aspect is addressed in the proxy variable framework by making a full use of the behavioral model of firm production, and we discuss this in detail in Section \ref{sec:identification}. We now consider the issues pertaining to the measurement of peer effects between firms.

The identification of learning and spillover effects on firm productivity in our model is based on several structural assumptions about the timing as well as the underlying form of peer interactions and network organization. 
To begin with, the productivity process in \eqref{eq:proddist_a_exp} is a dynamic analogue of a ``pure endogenous-effects model'' in Manski's nomenclature. It postulates that the cross-firm peer interactions occur only through the outcomes (i.e., $\omega$) whereby each firm's productivity is affected by the mean productivity of the peers in its reference group. As noted in Remark \ref{rem:omegalag_vs_Glag}, we effectively assume away the ``contextual effects'' of the peers' productivity modifiers and, in doing so, address the first of two \citeauthor{manski1993}'s (1993) unidentification results about the indistinguishability of endogenous and exogenous peer effects.\footnote{His second result is about the difficulty to distinguish ``real'' peer interactions through observables from the unobservable ``correlated effects;'' more on this later.} The latter issue becomes moot because in the absence of contextual effects of $\{G_{j,t-1};\ j(\ne i)=1,\dots,n\}$ on $\omega_{it}$ our model postulates a single channel of cross-peer dependence. Appendix \ref{sec:appx_add_ident} discusses how our setup may be augmented to allow for such contextual effects.

The evolution process in \eqref{eq:proddist_a_exp} also implicitly assumes that learning occurs with a delay which is why the dependence of $\omega_{it}$ on both its own productivity-modifying controls and peers' productivity is lagged, implying that the improvements in firm productivity take a period to materialize. Furthermore, in $\mathbb{E}[\zeta_{it} | \mathcal{I}_{i,t-1}]=0$ we assume that firms do not experience changes in their location and/or productivity modifiers in light of expected \textit{future} innovations in productivity. This timing assumption about the arrival of $\zeta_{it}$ renders both the lagged $G_{i,t-1}$ and a set of spatially proximate peers $\mathcal{L}(i,t-1)$ at time $t-1$ that defines the peer weights $\{s_{ij,t-1};\ j(\ne i)=1,\dots,n\}$ predetermined (weakly exogenous) with respect to the firm $i$'s productivity innovation at time $t$, which helps identify both the learning and spillover effects on firm productivity. 

When it comes to internal learning effects (via own $G_{i,t-1}$), such a timing assumption is quite common in the productivity literature \citep[e.g., see][]{vanbiesebroeck2005, deloecker2013,dj2013, malikovetal2015ijio}. More specifically, $\mathbb{E}[\zeta_{it} | \mathcal{I}_{i,t-1}]=0$ rules out the firm's ability to systematically predict future productivity shocks. Instead, the Markovian process in \eqref{eq:proddist_a_exp} states that the firm anticipates the effect of its $G$ productivity modifier on $\omega_{it}$ in period $t$ when adjusting the former in period $t-1$, and the conditional mean $\mathbb{E}\left[\omega_{it} | \omega_{i,t-1}, G_{i,t-1},\right.$ $\left.\sum_{j(\ne i)}s_{ij,t-1}\omega_{j,t-1}\right]$ is what captures that \textit{expected} productivity. But the \textit{actual} firm productivity at time $t$ also includes a random innovation $\zeta_{it}$. Essentially, the conditional-expectation-function error $\zeta_{it}$ represents unpredictable uncertainty that is naturally associated with productivity-modifying activities (new R\&D investments, entering export markets or attracting new foreign investors) such as chance in discovery, success in implementation, etc. This productivity innovation $\zeta_{it}$ is realized after $G_{i,t-1}$ is fully determined.\footnote{Depending on the source of learning, it may be possible to reasonably relax the assumption of a delayed learning effect of $G_{it}$ on firm productivity. See discussion in Appendix \ref{sec:appx_add_ident}.} 

In our paper, we also extend this timing assumption to external cross-firm learning via spillovers, which yields mean-orthogonality of the spatiotemporal ``lag'' of peers' productivities $\sum_{j(\ne i)}s_{ij,t-1}\omega_{j,t-1}$ and the innovation $\zeta_{it}$. That is, the assumed is weak exogeneity of the location-dependent peer weights $\{s_{ij,t-1}\}$, according to which firms do not relocate in anticipation of \textit{future} productivity shocks because such shocks are purely random. This rules out endogeneity of the firm's peer network in period $t-1$ with respect to the productivity shock $\zeta_{it}$ it experiences at time $t$. The plausibility of this is further buttressed by the fact that firm \textit{re}location in most industries (e.g., agriculture, manufacturing,  utilities) is highly, if not prohibitively, costly. In fact, our assumption about the weak exogeneity of group membership is not as strong as the standard assumption of fixed (non-random) networks commonly made in the (empirical) social-effects or spatial literature.

Note that our timing assumption about learning and spillover effects does \textit{not} rule out a \textit{contemporaneous} correlation between firm productivity and its productivity modifiers or even the location. That is, we do not assume that $\mathbb{E}[\zeta_{it} | \mathcal{I}_{it}]=0$. Consequently, firms are permitted to endogenously update their  $G_{it}$ as well as to change their locations based on the (observable by firms) period $t$ level of their productivity $\omega_{it}$. For instance, in the presence of inbound FDI opportunities that can help improve a domestic firm's productivity (when $G_{it}$ measures the firm's exposure to investors from abroad), the more productive firms are more likely to be attractive for investors, and the corresponding non-zero $\text{Cov}[G_{it},\omega_{it}]$ is well within our framework. 

An important implication of our structural assumption about $\mathbb{E}[ \zeta_{it} | \sum_{j(\ne i)}s_{ij,t-1}\omega_{j,t-1}] = 0$ is that the innovation in the productivity evolution process \eqref{eq:proddist_a_exp} does \textit{not} contain any unobservable ``correlated effects''  at the reference group level\textemdash to borrow Manski's terminology\textemdash the presence of which can complicate, if not hinder, the identification of peer effects occurring through the group mean productivity $\sum_{j(\ne i)}s_{ij,t-1}\omega_{j,t-1}$. Effectively, we attribute all cross-firm dependence in productivity to the within-group dependence of each firm's underlying productivity on that of its peers as opposed to the tendency of all group firm-members to see their productivities evolve in a similar fashion due to the influence of common group unobservables such as shared locational/institutional environments. This is an admittedly strong but fairly common working assumption in the literature, given the well-known challenges in tackling group-level unobservables in network models \citep[for an excellent review, see][]{blumeetal2011}. Our no-group-effects assumption echoes the existing studies of R\&D/FDI/export spillovers and the productivity literature more broadly, and we maintain it to maximize comparability with the commonly used methodologies. Having said that, this assumption can be relaxed\textemdash we do so in our robustness checks\textemdash if we restrict the group-level unobservables to be time-invariant \`a la \citet{grahamhahn2005} and \citet{bramoulleetal2009}.

Fundamentally, the potential threats to identification of the spillover effects posed by the correlated group effects can otherwise be cast as a spatial selection/sorting problem, whereby more productive firms may be \textit{ex ante} sorting into the what-then-become high productivity locations. Under this scenario, when we compare the firm to its spatial peers, we may mistakenly attribute any future productivity improvements to spillovers from the peers (i.e., agglomeration), while in actuality it merely reflects the underlying propensity of \textit{all} firms in this location to be more productive and, consequently, more apt at improving their productivity. While there has recently been notable progress in formalizing and understanding these coincident phenomena theoretically \citep[e.g.,][]{behrensetal2014,gaubert2018}, disentangling firm sorting and agglomeration remains a non-trivial task empirically.\footnote{Urban economics literature also distinguishes the third endogenous process usually referred to as the ``selection'' which differs from sorting in that it occurs \textit{ex post} after the firms had self-sorted into locations and which determines their continuing survival. We abstract away from this low-productivity-driven attrition issue in the light of the growing empirical evidence suggesting that it explains none of spatial productivity differences which, in contrast, are mainly driven by agglomeration economies \citep[see][]{combesetal2012}. Relatedly, the firm attrition out of the sample has also become commonly accepted as a \textit{practical} non-issue in the productivity literature so long as the data are kept unbalanced. For instance, \citet[][p.324]{lp2003} write: ``The original work by Olley and Pakes devoted significant effort to highlighting the importance of not using an artificially balanced sample (and the selection issues that arise with the balanced sample). They also show once they move to the unbalanced panel, their selection correction does not change their results.''} However, by including the firm's \textit{own} lagged productivity in the autoregressive $\omega_{it}$ process in \eqref{eq:proddist_a_exp}, we are able (at least to some extent) to account for this potential self-sorting because sorting into locations is heavily influenced by the firm's own productivity (oftentimes stylized as the ``talent'' or ``efficiency'' in theoretical models). That is, the spillover effect $SP_{it}$ on future firm productivity in our model is measured after partialling out the contribution of its own productivity. Incidentally, \citet{deloecker2013} argues the same in the context of export-based learning and self-selection of exporters.

We maintain the \textit{i.i.d.}~assumption about the random transitory shock $\eta_{it}$, from where it follows that $\mathbb{E}[\eta_{it} | \mathcal{I}_{it}] = \mathbb{E}[\eta_{it}] = 0$ with the mean normalized to zero. The latter implies that the shock $\eta_{it}$ is observable to firms in period $t$ only \textit{ex post} after all production decisions.

% ------------------------------------------------------------------------------------------
% ------------------------------------------------------------------------------------------

\section{A System Approach to Identification via Proxy Variables}\label{sec:identification}

Logging the production function in \eqref{eq:prodfn} and making use of the Markovian nature of $\omega_{it}$ from \eqref{eq:proddist_a_exp}, we obtain
\begin{equation}\label{eq:prodfn_gross_logs_2}
	y_{it} = \beta_Kk_{it}+\beta_Ll_{it}+\beta_Mm_{it} + h\left(\omega_{i,t-1}, G_{i,t-1}, \sum_{j(\ne i)}s_{ij,t-1}\omega_{j,t-1}\right) + \zeta_{it} + \eta_{it} ,
\end{equation}
where the lower-case variables denote the logs of the respective upper-case variables, and $h[\cdot]\equiv\mathbb{E}[\omega_{it} |\cdot]$ is some unknown function. Under our structural assumptions about firm behavior, all right-hand-side covariates in \eqref{eq:prodfn_gross_logs_2} are predetermined and thus mean-independent of $\zeta_{it} + \eta_{it}$, except for the freely varying input $m_{it}$ that the firm chooses in time period $t$ conditional on $\omega_{it}$ (among other state variables including quasi-fixed inputs) thereby making it a function of $\zeta_{it}$. That is, the materials variable is endogenous.

To consistently estimate \eqref{eq:prodfn_gross_logs_2}, we first need to address the latency of firm productivity $\omega_{it}$. A widely popular solution to this problem in the literature is to adopt a proxy variable approach \`a la \citet{lp2003} whereby unobservable $\omega_{it}$ is proxied by inverting the firm's conditional demand for an observable static input $m_{it}$. However, \citet{gnr2013} show that identification generally fails under such a standard estimation procedure due to the lack of a valid instrument for the endogenous $m_{it}$ despite the abundance of predetermined higher-order lags of inputs. Therefore, the production function remains unidentified in the flexible input. To solve this problem, they suggest exploiting a structural link between the production function and the firm's (static) optimality condition. In what follows, we build on this idea which we adapt in the spirit of \citet{dj2013}, whereby we explicitly make use of the assumed functional form of the production function to streamline identification of the material elasticity and to ease computational burden of estimation (also see Remark \ref{remark:funcform}).

We first focus on the identification of the production function in its flexible input $M_{it}$. Specifically, given the Cobb-Douglas form, we seek to identify the material elasticity parameter $\beta_M$. To do so, we consider an equation for the firm's first-order condition with respect to $M_{it}$. Since it is a static input, the firm's optimal choice of $M_{it}$ can be modeled as the restricted expected profit-maximization problem\footnote{Under the risk neutrality of firms.} subject to the (already) optimal allocation of quasi-fixed inputs:
\begin{align}\label{eq:profitmax}
\max_{M_{it}}\ P_{t}^Y K_{it}^{\beta_K}L_{it}^{\beta_L}M_{it}^{\beta_M} \exp\{\omega_{it}\}\theta  - P_{t}^M M_{it} ,
\end{align}
where $P_{t}^Y$ and $P_{t}^M$ are respectively the output and material prices that, under the commonly invoked assumption of perfect competition, need not vary across firms; and $\theta\equiv\mathbb{E}[\exp\{\eta_{it}\}|\ \mathcal{I}_{it}]$. The first-order condition is given by
\begin{equation}\label{eq:profitmax_foc}
\beta_M P_{t}^Y K_{it}^{\beta_K}L_{it}^{\beta_L}M_{it}^{\beta_M-1} \exp\{\omega_{it}\} \theta = P_{t}^M ,
\end{equation}
which can be transformed via dividing it by the production function in \eqref{eq:prodfn} to obtain the following stochastic material share equation (in logs):
\begin{equation}\label{eq:fst}
\ln V_{it} = \ln (\beta_M\theta) - \eta_{it} ,
\end{equation}
where $V_{it} \equiv P_{t}^M M_{it}/(P_{t}^Y Y_{it})$ is the nominal share of material costs in total revenue. This share is readily observable in the data, and the construction thereof does not require firm-level prices. 

Intuitively, equation \eqref{eq:fst} says that \textit{unobservable} material elasticity of the production function $\beta_M$ can be identified from \textit{observable} material share $V_{it}$ because the two must be equal on average (in logs) to maximize profits. Specifically, it identifies $\beta_M\times\theta$ (and the random productivity residual $\eta_{it}$) based on $\mathbb{E}[\eta_{it}]=0$:
\begin{equation}\label{eq:fst_ident_theta}
\ln(\beta_M\theta) = \mathbb{E}[\ln V_{it}].
\end{equation}

To identify the material elasticity $\beta_M$ net of constant $\theta$, recognize that $\theta$ can be identified via $\theta =\mathbb{E}[ \exp\{\eta_{it}\}]= \mathbb{E}\left[ \exp\{ \ln (\beta_M\theta)-\ln V_{it}\} \right]=\mathbb{E}\left[ \exp\{ \mathbb{E}[\ln V_{it}]-\ln V_{it}\} \right]$. Then, we have that
\begin{equation}\label{eq:fst_ident}
\beta_M = \exp\left\{ \mathbb{E}[\ln V_{it}] \right\}\big/ 
\mathbb{E}\left[ \exp\{ \mathbb{E}[\ln V_{it}]-\ln V_{it}\} \right].
\end{equation}

With $\beta_M$ identified from \eqref{eq:fst_ident}, we have thus identified the production function in the dimension of its endogenous freely varying input $M_{it}$ thereby effectively circumventing \citeauthor{gnr2013}'s (2017) critique. To see the latter, we rewrite \eqref{eq:prodfn_gross_logs_2} as follows: 
\begin{equation}\label{eq:prodfn_gross_logs_3}
y_{it}^* = \beta_Kk_{it}+\beta_Ll_{it} + h\left(\omega_{i,t-1}, G_{i,t-1}, \sum_{j(\ne i)}s_{ij,t-1}\omega_{j,t-1}\right) + \zeta_{it} + \eta_{it} ,
\end{equation}
where $y_{it}^*\equiv y_{it} - \beta_M m_{it}$ is already identified and thus observable, and our model in \eqref{eq:prodfn_gross_logs_3} no longer contains endogenous variables needing instrumentation.

To identify the remaining parameters of the production function $(\beta_K,\beta_L)'$ as well as latent firm productivity $\omega_{it}$ in \eqref{eq:prodfn_gross_logs_3}, we make use of the \textit{known}  parametric form of the conditional material demand function $M_{it}=\mathbb{M}(\omega_{it},K_{it},L_{it},P_{t}^Y,P_{t}^M)$ implied by the first-order condition in \eqref{eq:profitmax_foc} which we invert for $\omega_{it}$. Under our standard assumptions about firm behavior and regularity conditions on the production function, $\mathbb{M}(\cdot)| M_{it}>0$ must be strictly monotonic in $\omega_{it}$ for any given $(K_{it},L_{it},P_{t}^Y,P_{t}^M)$, and hence we can invert $\mathbb{M}(\cdot)$ to control for unobserved persistent productivity via $\omega_{it}= \mathbb{M}^{-1}(M_{it},K_{it}, L_{it},P_{t}^Y,P_{t}^M)$. Specifically, substituting for $\omega_{i,t-1}$ and $\omega_{j,t-1}$ using the inverted material function derived analytically from \eqref{eq:profitmax_foc}, from \eqref{eq:prodfn_gross_logs_3} we get
\begin{align}\label{eq:sst}
y_{it}^* =&\ 
\beta_Kk_{it}+\beta_Ll_{it} + h\left(\omega_{i,t-1}^*\left(\beta_K,\beta_L\right), 
G_{i,t-1}, 
\sum_{j(\ne i)}s_{ij,t-1} \omega_{j,t-1}^*\left(\beta_K,\beta_L\right)\right) + \zeta_{it} + \eta_{it} ,
\end{align}
where 
\begin{align}\label{eq:omega_starr}
\omega_{it}^*\left(\beta_K,\beta_L\right) = \Big[(1-\beta_M)m_{it}-\ln(\beta_M\theta) -\ln(P_t^Y/P_t^M)\Big]- \beta_Kk_{it}-\beta_Ll_{it}\quad \forall i,t 
\end{align}
is the inverted material demand function in which the bracketed component is already observable and therefore the only remaining unknown parameters in it are $(\beta_K,\beta_L)'$. All right-hand-side covariates in the semiparametric model \eqref{eq:sst} are weakly exogenous and thus self-instrument. The model is thus identified on the basis of  
\begin{equation}\label{eq:sst_ident}
\mathbb{E}\left[ \zeta_{it} + \eta_{it} \Bigg| k_{it},l_{it},k_{i,t-1},l_{i,t-1},m_{i,t-1},G_{i,t-1}, \sum_{j\ne i}s_{ij,t-1}k_{j,t-1},\sum_{j\ne i}s_{ij,t-1}l_{j,t-1},\sum_{j\ne i}s_{ij,t-1}m_{j,t-1}\right] = 0  ,
\end{equation}
where we have made explicit use of the variables entering the proxy function $\omega_{it}^*\left(\beta_K,\beta_L\right)$. 

The appearance of group averages of the peers' predetermined inputs in \eqref{eq:sst_ident} is akin to the idea of instrumenting the endogenous group mean of an outcome with the exogenous group mean characteristics, which is a common identification strategy in both the social-effects and spatial econometrics literature \citep[e.g., see][]{lesagepace2009,bramoulleetal2009}. The critical distinction here is that, in our case, the ``group mean of an outcome'' $\sum_{j\ne i}s_{ij,t-1}\omega_{j,t-1}$ is \textit{not} endogenous with respect to $\zeta_{it} + \eta_{it}$ and therefore needs no instrumentation. In contrast, our use of the ``group mean characteristics'' $\big(\sum_{j\ne i}s_{ij,t-1}k_{j,t-1},\sum_{j\ne i}s_{ij,t-1}l_{j,t-1},\sum_{j\ne i}s_{ij,t-1}m_{j,t-1}\big)'$ is effectively in their proxy-variable capacity given latency of $\omega_{j,t-1}$.

\begin{remark}\normalfont\label{remark:funcform} Following the steps of \citet{dj2013}, our approach fully embraces the assumed parametric specification of the firm's production function by explicitly utilizing the known functional form of the first-order condition for materials and the inverse conditional input demand function that it implies. By doing so, we circumvent the need to integrate the estimated material elasticity function at \textit{each} observation in order to recover the unknown production function required by \citeauthor{gnr2013}'s (2017) nonparametric methodology. Importantly, by relying on parameter restrictions between the production function and inverted material demand function in \eqref{eq:sst}, we do not have to rely on nonparametric methods to estimate the unknown proxy function for $\omega$ that appears inside the also unknown $h(\cdot)$ function. Otherwise, identification of \eqref{eq:sst} would have been complicated by the presence of a nonparametric $\mathbb{M}^{-1}(\cdot)$ function (evaluated at multiple data points\footnote{That is, evaluated at $(m_{i,t-1},k_{i,t-1},l_{i,t-1})$ to proxy for $\omega_{i,t-1}$ as well as at $(m_{j,t-1},k_{j,t-1},l_{j,t-1})\ \forall\ j$ to proxy for $\omega_{j,t-1}$ entering the spillover-capturing peer group average.}) inside another nonparametric function $h(\cdot)$. Our parametric inversion of $\mathbb{M}^{-1}(\cdot)$  yields a much simpler semiparametric estimator. 
\end{remark}

\begin{remark} \normalfont\label{remark:acfcritique}  Our model is also robust to \citeauthor{acf2015}'s (2015) critique that focuses on the potential inability of structural proxy variable estimators to separably identify the production function and productivity proxy. This issue arises in the wake of perfect functional dependence between variable inputs appearing both inside the unknown production function and productivity proxy function. Our second-stage equation \eqref{eq:sst} does not suffer from such a problem because it contains no (endogenous) variable input on the right-hand side, the corresponding parameter of which has already been identified from the share equation in the first stage.
\end{remark}

Lastly, with all parameters of the production function $(\beta_K,\beta_L,\beta_m)'$ and the transitory productivity shock ${\eta}_{it}$ successfully identified in the two stages, we readily identify latent firm productivity $\omega_{it}$ from the production function in logs: $\omega_{it}=y_{it}- \beta_Kk_{it}-\beta_Ll_{it}-\beta_Mm_{it}-\eta_{it}$.

% ------------------------------------------------------------------------------------------
% ------------------------------------------------------------------------------------------

\section{Estimation Procedure}
\label{sec:estimation}

We implement our identification strategy in two stages. In the first stage, we estimate the material elasticity of the production function. Based on \eqref{eq:fst_ident}, the consistent estimator of $\beta_M$ is
\begin{equation}\label{eq:fst_est}
\widehat{\beta}_M = \exp\Big\{ \frac{1}{N}\sum_i\sum_t \ln V_{it} \Big\}\Big/ \Big[\frac{1}{N}\sum_i\sum_t \exp\Big\{ \Big(\frac{1}{N}\sum_i\sum_t \ln V_{it}\Big) - \ln V_{it} \Big\} \Big],
\end{equation}
where $N$ is the total number of observations, which equals $nT$ in the case of a balanced panel.

We then estimate $y_{it}^*$ via $\widehat{y}_{it}^* = y_{it} - \widehat{\beta}_M m_{it}$ and also construct ``partial'' estimates of the productivity proxy function $\omega_{it}^*\left(\beta_K,\beta_L\right)$ in \eqref{eq:omega_starr} as
\begin{align*}
\widehat{\omega}_{it}^*\left(\beta_K,\beta_L\right) = \underbrace{\Big[(1-\widehat{\beta}_M)m_{it}-\ln(\widehat{\beta_M\theta}) -\ln(P_t^Y/P_t^M)\Big]}_{\textstyle \widehat{\varkappa}_{it}}- \beta_Kk_{it}-\beta_Ll_{it},
\end{align*}	
where $\ln(\widehat{\beta_M\theta})=\frac{1}{N}\sum_i\sum_t \ln V_{it}$ on the basis of \eqref{eq:fst_ident_theta}. Note that $\widehat{\omega}_{it}^*\left(\beta_K,\beta_L\right)$ still contains two unknowns which enter the function linearly: $(\beta_K,\beta_L)'$. For convenience, let the already identified portion of productivity be denoted by $\widehat{\varkappa}_{it}= (1-\widehat{\beta}_M)m_{it}-\ln(\widehat{\beta_M\theta}) -\ln(P_t^Y/P_t^M)$.

With $\widehat{y}_{it}^*$ and $\widehat{\omega}_{it}^*\left(\beta_K,\beta_L\right)$ from the first stage in hand, we proceed to the second-stage estimation of \eqref{eq:sst}, where we approximate unknown function $h(\cdot)$ using polynomial sieves. Specifically, recognize that $\widehat{\omega}_{it}^*\left(\beta_K,\beta_L\right)=\widehat{\varkappa}_{it}-\beta_Kk_{it}-\beta_Ll_{it}$ and let $\widehat{\mathbf{z}}_{i,t-1}(\boldsymbol{\beta})= ([\widehat{\varkappa}_{i,t-1}- \beta_Kk_{i,t-1}-\beta_Ll_{i,t-1}],$ $G_{i,t-1}, \sum_{j(\ne i)}s_{ij,t-1}[\widehat{\varkappa}_{j,t-1}- \beta_Kk_{j,t-1}-\beta_Ll_{j,t-1}] )'$ be a $3\times 1$ vector with $\boldsymbol{\beta}=(\beta_K,\beta_L)'$. Then, for each $\widehat{\boldsymbol{z}}(\boldsymbol{\beta})$, we approximate the unknown function $h\left(\widehat{\boldsymbol{z}}(\boldsymbol{\beta})\right)$ in \eqref{eq:sst} as follows:
\begin{equation}\label{eq:happrox}
h\left(\widehat{\boldsymbol{z}}(\boldsymbol{\beta})\right) \approx \mathcal{A}_{L_{n}}\left( \widehat{\boldsymbol{z}}(\boldsymbol{\beta})\right) ^{\prime }\boldsymbol{\gamma} ,
\end{equation}
where $\mathcal{A}_{L_{n}}\left(\cdot\right) =\left( A_{1}\left( \cdot\right) ,\dots ,A_{L_{n}}\left( \cdot\right) \right) ^{\prime }$ is an $L_{n} \times 1$ vector of known basis functions of $\widehat{\boldsymbol{z}}(\boldsymbol{\beta})$ including a vector of ones, $\boldsymbol{\gamma}$ is a conformable vector of parameters, and $L_n \to\infty$ slowly with $n$.

Given the orthogonality conditions in \eqref{eq:sst_ident}, we estimate $\boldsymbol{\beta}$ and $\boldsymbol{\gamma}$ via nonparametric nonlinear least squares. Letting $\mathbf{x}_{it}=(k_{it},l_{it})'$, the parameter estimators are given by
\begin{align}\label{eq:sst_est}
	\left(\widehat{\beta}_K,\widehat{\beta}_L, \widehat{\boldsymbol{\gamma}}'\right)'=
	\argmin_{\boldsymbol{\beta},\boldsymbol{\gamma}}\ \sum_{i}\sum_t \left[\widehat{y}_{it}^*- \mathbf{x}_{it}'\boldsymbol{\beta} - \mathcal{A}_{L_{n}}\left( \widehat{\mathbf{z}}_{i,t-1}(\boldsymbol{\beta})\right) ^{\prime }\boldsymbol{\gamma}\right]^2,
\end{align}
where the minimand is the sum of squared errors corresponding to our sieve estimator of \eqref{eq:sst}.

Using the already estimated $(\widehat{\beta}_K,\widehat{\beta}_L,\widehat{\beta}_M)'$ and $\widehat{\boldsymbol\gamma}$, we then readily have the estimators for our primary estimands of interest: $\widehat{DL}_{it}\equiv\partial \widehat{h}_{it}(\cdot)/\partial G_{i,t-1}$, $\widehat{SP}_{it}\equiv\partial \widehat{h}_{it}(\cdot)/\partial \sum_{j(\ne i)}s_{ij,t-1}\widehat{\omega}_{j,t-1}$ and $TIL_{it} = \widehat{SP}_{it} \times \sum_{j(\ne i)} s_{ij,t-1} \widehat{DL}_{j,t-1}$ respectively measuring the direct learning, cross-firm spillover and total indirect learning effects, where $\widehat{h}_{it}(\cdot) = \mathcal{A}_{L_{n}}\big( \widehat{\mathbf{z}}_{it}(\widehat{\boldsymbol{\beta}})\big)^{\prime} \widehat{\boldsymbol{\gamma}}$ and $\widehat{\omega}_{j,t-1}= \widehat{\varkappa}_{j,t-1} - \widehat{\beta}_Kk_{j,t-1}- \widehat{\beta}_Ll_{j,t-1}$. Lastly, the estimator of latent firm productivity is $\widehat{\omega}_{it}=y_{it}- \widehat\beta_Kk_{it}-\widehat\beta_Ll_{it}-\widehat\beta_Mm_{it}-\widehat{\eta}_{it}$, 
where $\widehat{\eta}_{it}=\ln (\widehat{\beta_M\theta}) - \ln V_{it}$ from the first stage.

For the limit results, our sequential estimation methodology can be recast as a moment-based semiparametric sieve M-estimation problem. Specifically, letting
\begin{equation*}
	\mathbf{z}_{i,t-1}\left(\beta_M,\beta_K,\beta_L,\theta\right)= 
	\begin{bmatrix}
		(1-\beta_M)m_{it-1}-\ln(\beta_M\theta) -\ln(P_{t-1}^Y/P_{t-1}^M) - \beta_Kk_{it-1}-\beta_Ll_{it-1} \\
		G_{it-1} \\
		\sum_{j(\ne i)}s_{ijt-1}\left[(1-\beta_M)m_{jt-1}-\ln(\beta_M\theta) -\ln(P_{t-1}^Y/P_{t-1}^M)- \beta_Kk_{jt-1}-\beta_Ll_{jt-1}\right]
	\end{bmatrix}
\end{equation*} 
and $r_{it}\left(\beta_M,\beta_K,\beta_L,\theta\right)=y_{it} - \beta_M m_{it}- \beta_Kk_{it}-\beta_Ll_{it} - \mathcal{A}_{L_{n}}\big( \mathbf{z}_{i,t-1}\left(\beta_M,\beta_K,\beta_L,\theta\right)\big)'\boldsymbol{\gamma}$, we can rewrite our two estimation stages in the form of the following multiple-equation moment restrictions:
\begin{equation}\label{eq:msys}
	\mathbb{E} 
	\begin{bmatrix}
		\ln V_{it} - \ln(\beta_M\theta) \\
		\exp\left\{ \ln(\beta_M\theta) - \ln V_{it} \right\} - \theta  \\
		r_{it}\left(\beta_M,\beta_K,\beta_L,\theta\right)
		\begin{bmatrix} \partial r_{it}\left(\beta_M,\beta_K,\beta_L,\theta\right)\big/\partial \left(\beta_K,\beta_L\right)'  \\ -\mathcal{A}_{L_{n}}\big( \mathbf{z}_{i,t-1}\left(\beta_M,\beta_K,\beta_L,\theta\right)\big) \end{bmatrix} 
	\end{bmatrix} = \mathbf{0}_{4+L_n},
\end{equation}
consisting of two blocks, where the first two moments correspond to the estimator of the material elasticity (first block) and the remaining orthogonality conditions correspond to the nonlinear sieve least-squares estimation of the proxied production function and productivity in \eqref{eq:sst_est}. 

In the above, $\mathcal{A}_{L_{n}}(\cdot)$ is a sieve approximation of the unknown infinite-dimensional nonparametric function $h(\cdot)$, and $\left(\beta_M,\beta_K,\beta_L,\theta\right)'$ are the unknown parameters of fixed dimension. Thus, our estimator falls within \citeauthor{aichen2003}'s (2003) general framework for the minimum distance estimation based on the conditional moment restrictions of a generic form $\mathbb{E}[\rho(X,\delta_0,g_0(\cdot)) |Z]=0$, where $X$ and $Z$ are data and $\rho(\cdot)$ is a vector of ``residual functions'' with finite-dimensional unknown parameters $\delta$ and infinite-dimensional unknown functions $g$. The large-sample limit results from \citet{aichen2003} and \citet{chenetal2003} therefore extend to our two-step estimator. 
Inference can be asymptotic or via bootstrap; we discuss both in detail in Appendix \ref{sec:inference}.

% ------------------------------------------------------------------------------------------
% ------------------------------------------------------------------------------------------

\section{Simulations}
\label{sec:simulations}

We conduct a set of Monte Carlo experiments.
%First, we evaluate the performance of our proposed estimator and demonstrate its ability to successfully identify firm productivity and, importantly, the cross-firm spillovers therein. Second, we use the simulations to show that estimating spillovers via an internally inconsistent two-step procedure using traditional proxy methods which ignore the spillover-induced cross-sectional dependence in firm productivity, as customarily done in the literature, can lead to spurious and misleading results due to the endogeneity-inducing misspecification in the measurement of productivity in the first step.
Our data generating process draws from those used by \citet{griecoetal2016} and \citet{gnr2013}. Specifically, we consider a balanced panel of $n=\{100,200,400\}$ firms operating during $T=10$ periods.\footnote{We have also experimented with 5 and 50 periods. The results are qualitatively unchanged.} Each panel is simulated 1,000 times. To simplify matters, we dispense with labor and consider the production process only with the quasi-fixed dynamic $K_{it}$ and freely-varying static $M_{it}$. The production technology is
\begin{equation}
Y_{it} = K_{it}^{\beta_K}M_{it}^{\beta_M}\exp\{\omega_{it}+\eta_{it}\},
\end{equation}
where we set $\beta_K=0.25$ and $\beta_M=0.65$, and the noise $\eta_{it}\sim\text{i.i.d.}\ \mathbb{N}(0,\sigma_{\eta}^2)$ with $\sigma_{\eta}=\sqrt{0.07}$.

The firm's capital is set to evolve according to $K_{it}= I_{i,t-1}+(1-\delta_i)K_{i,t-1}$, with the firm-specific depreciation rates $\delta_i\in \{0.05,0.075,0.10,0.125,0.15\}$ distributed uniformly across $i$. The initial levels of capital $K_{i0}$ is drawn from \textit{i.i.d.}~$\mathbb{U}(10,200)$. The investment function takes the following form: $I_{i,t-1}= K_{i,t-1}^{\alpha_1} \exp\{\alpha_2\omega_{it-1}\}$, where $\alpha_1=0.8$ and $\alpha_2=0.1$.

The materials $M_{it}$ series is generated solving the firm's restricted expected profit maximization problem along the lines of \eqref{eq:profitmax}. The conditional demand for $M_{it}$ is given by
\begin{align}
M_{it}&= \argmax_{\mathcal{M}_{it}}\ \Big\{ P_{t}^Y K_{it}^{\beta_K}\mathcal{M}_{it}^{\beta_M} \exp\{\omega_{it}\}\theta  - P_{t}^M \mathcal{M}_{it} \Big\}  
=\left( \beta_M K_{it}^{\beta_K}\exp\{\omega_{it}\}\right)^{1/(1-\beta_M)},
\end{align}
where, in the second equality, we have normalized $P_t^M=\theta\ \forall\ t$ and have assumed no temporal variation in output prices: $P^{Y}_{t}=1$ for all $t$.

We assume that the firm's productivity modifier $G_{it}$ is autoregressively persistent. More specifically, we consider two laws of motion for $G_{it}$: (\textsl{a}) an exogenous autoregressive process whereby $G_{it}=\gamma_0+\gamma_1G_{i,t-1}+\epsilon_{it}$ and (\textsl{b}) a controlled autoregressive process, contemporaneously conditional on the firm's latent productivity: $G_{it}= \gamma_0+\gamma_1G_{i,t-1}+\gamma_2\omega_{it}+\epsilon_{it}$, where $\gamma_0=0.01$, $\gamma_1=0.6$, $\gamma_2=0.3$ and $\epsilon_{it}\sim\text{i.i.d.}\ \mathbb{N}(0,\sigma_{\epsilon}^2)$ with $\sigma_{\epsilon}=0.1$. Of the two processes, the second one assumes that more productive firms engage in higher levels of the productivity-modifying activities. The process (\textsl{b}) permits firms to endogenously update their $G_{it}$ based on the (observable by them) period $t$ level of their productivity $\omega_{it}$. For example, if $G_{it}$ measures the firm's exposure to investors from abroad, this accommodates the scenario when foreign investors choose to invest in more productive domestic firms in the first place.

We let firm productivity $\omega_{it}$ be a linear spatiotemporal first-order autoregressive process:
\begin{align}\label{eq:omega_sim}
	\omega_{it}=\rho_0+\rho_1\omega_{i,t-1}+\rho_2\sum_{j(\ne i)}s_{ij,t-1}\omega_{j,t-1}+\rho_3G_{i,t-1}+\zeta_{it},
\end{align}
where, unless stated otherwise, we set $\rho_0=0.2$, $\rho_1=0.55$, $\rho_2=0.4$ and $\rho_3=0.5$.  The innovation is generated as $\zeta_{it}\sim\text{i.i.d.}\ \mathbb{N}(0,\sigma_{\zeta}^2)$ with $\sigma_{\zeta}=0.2$. The initial level of productivity $\omega_{i0}\sim \text{i.i.d.}~\mathbb{U}(1,3)$ over $i$. In Appendix \ref{sec:appendix_sim_nl}, we also present the results for a \textit{nonlinear} specification for $\omega_{it}$.
	
To keep matters simple, we consider one common spatial region for all firms and assume that all firms belong to the same industry. Hence, cardinality of the set $\mathcal{L}(i,t)$ is the same across all $i$ and equals $n-1$. The peer weights $\{s_{ijt};\ j(\ne i)\}$ are constructed according to \eqref{eq:weight1} and, given the setup, are equal to $1/(n-1)$ for all firms and time periods. 
	
%------------------------------------------------------------------------------------------
\paragraph{Proposed Methodology.} First, we evaluate the performance of our proposed estimator with the focus on its ability to successfully identify productivity spillovers across firms. For each combination of the $G$ and $\omega$ processes, we consider the following three DGP scenarios: 
(\textbf{{i}}) a general case scenario in which firm productivity is modified via both the direct learning ($DL_{it}\ne0$) and cross-firm spillovers ($SP_{it}\ne0$);
(\textbf{{ii}}) a special case scenario in which we assume no direct learning ($DL_{it}=0$ globally) in order to focus our attention exclusively on the agglomeration-driven learning via spillovers ($SP_{it}\ne0$); 
(\textbf{{iii}}) an even more special case scenario in which firm productivity evolves exogenously (both $DL_{it}=0$ and $SP_{it}=0$ globally) as traditionally assumed in the proxy variable production function estimation literature. The special case scenarios are implemented by setting the appropriate coefficients in the productivity process \eqref{eq:omega_sim} to zero.

We estimate the model via the two-stage estimation algorithm outlined in Section \ref{sec:estimation}, where we approximate unknown $h(\cdot)$ using second-degree polynomial sieves. Table \ref{tab:sim1s} reports the simulation results for our proposed estimator, when $G_{it}$ evolves exogenously [top panel] and following an $\omega_{it}$-dependent controlled process [bottom panel]. Each of these two panels includes the results from the three different scenarios.  Reported are the mean, root mean squared error (RMSE) and mean absolute error (MAE) of the fixed-parameter $\beta_K$ estimates\footnote{We omit the results corresponding to the material elasticity $\beta_M$ from the first stage because the estimator yields very precise estimates of $\beta_M$ via \eqref{eq:fst_est} with the MSE and MAE being at least as small as $10^{-10}$ owing to the small sampling error induced by $\eta_{it}$ in our DGPs. We also experimented with much larger values of $\sigma_{\eta}$ with no significant changes to the qualitative results.} and the averages (across simulation iterations) of these metrics corresponding for each iteration computed using observation-specific nonparametric estimates of the autoregressive gradient $AR_{it}=\partial h(\cdot)/\partial \omega_{i,t-1}$, $DL_{it}=\partial h(\cdot)/\partial G_{i,t-1}$, $SP_{it}=\partial h(\cdot)/\partial \sum_{j\ne i}\omega_{j,t-1}$ and $TIL_{it}=SP_{it} \times \sum_{j(\ne i)} s_{ij,t-1} DL_{j,t-1}$.

The results in Table \ref{tab:sim1s} are encouraging and show that our methodology recovers the true parameters remarkably well, thereby lending strong support to the validity of our identification strategy. As expected of a consistent estimator, the estimation becomes more stable as $n$ grows. Same is the case when the productivity DGP in nonlinear (see Table \ref{tab:sim1g} in Appendix \ref{sec:appendix_sim_nl}).

%------------------------------------------------------------------------------------------
\paragraph{Alternative Procedures.} Next, to demonstrate the advantage of our internally consistent methodology, we inspect the performance of a widely used alternative procedure for estimating spillovers via a two-step approach. In this case, the unobserved firm productivity $\omega_{it}$ is first estimated via the standard proxy variable estimator (which assumes that the productivity process is an exogenous Markov chain and thus ignores spillovers) and then linearly regressed on some measure of the firm's exposure to its peers in the second step. As already discussed at length, such second-step regressions are inconsistent with the assumptions made in the first step because they contradictorily postulate the existence of cross-peer dependence which was assumed away when recovering firm productivity in the first place. Consequently, the productivity estimates (by means of the production function) obtained via such an approach are prone to biases due to the endogeneity-inducing misspecification of the productivity proxy. The empirical evidence of spillovers can thus be spurious. This is unsurprising because the unaccounted cross-sectional dependence is a hindrance to identification in general \citep[see][]{pesaran2006,bai2009}.

The second-step regressions used in spillovers literature have numerous variations but can be by and large categorized into two distinct types: those that measure the firm's exposure to spillovers from peers using the group means of characteristics which are said to facilitate such spillovers (FDI, R\&D, exports, etc.), and those that measure the firm's exposure to spillovers using the peer group mean of an outcome (that is, firm productivity). Essentially, the first type of regressions focuses on the ``contextual effects'' while the second type models cross-peer dependence via ``endogenous effects.'' Rarely do researchers allow for both effects at the same time. The first type is arguably the predominant choice in spillovers literature.  Such studies overwhelmingly estimate linear specifications, and virtually all omit the temporal lag of the firm's own productivity in the second-step analysis. 

We consider alternative methodologies with the second-step regressions of both these types. 
To facilitate a level-playing-field comparison between these and our models in the ability to identify spillovers, we specify the second-step regressions in lags. This is to ensure the maximal compatibility of the second-step regressions with the fashion in which learning and spillovers occur in the DGP. For concreteness, we run the following two second-step regressions:
\begin{align}
[\text{ALT1}]\qquad  \widehat{\omega}_{it} &= \alpha_{11} + \alpha_{12}\sum_{j(\ne i)}s_{ij,t-1}G_{j,t-2}+ \alpha_{13}G_{i,t-1}+e_{1,it},  \label{eq:alt1} \\
[\text{ALT2}]\qquad \widehat{\omega}_{it} &= \alpha_{21} + \alpha_{22}\sum_{j(\ne i)}s_{ij,t-1}\widehat{\omega}_{j,t-1}+ \alpha_{23}G_{i,t-1}+e_{2,it}, \label{eq:alt2}
\end{align}
using $\widehat{\omega}_{it}$ recovered in the first step via our semiparametric production function estimator but assuming an exogenous Markov process for productivity  $\omega_{it}=\mathbb{E}[\omega_{it}|\omega_{i,t-1}]+\zeta_{it}$.\footnote{Essentially, firm productivity here is estimated via the semiparametric adaptation of the original \citet{gnr2013} procedure modified to take advantage of the ``known'' parametric form of the production technology.} Here we also permit the $DL$ effects as oftentimes done in this literature. Because regressors in both alternative procedures in  \eqref{eq:alt1}--\eqref{eq:alt2} are all weakly exogenous per our DGP and the assumptions, these second-step regressions are estimates via least squares.

To be able to meaningfully analyze the performance of alternative models as well as to fairly compare them to our methodology (especially, in case of the popular ALT1 specification), we focus on the estimands that match in terms of their \textit{qualitative} interpretations. Instead of looking at specific parameters that may not always be directly comparable across the models and with the DGP, we consider the derived measures of $DL$, $SP$ and $TIL$ as appropriate/available. For instance, of the two alternative methodologies, only the ALT2 specification postulates cross-firm spillovers via the mean peer {productivity} as in our proposed conceptualization in Section \ref{sec:model} and the DGP. Therefore, $\alpha_{22}$ is essentially comparable to $\rho_2$ in the DGP: both measure the $SP$ effect. This is however not the case with the ALT1 specification which only models the contextual effect. Hence, we cannot contrast $\alpha_{12}$ to the true $\rho_2$ value in the DGP. Having said that, $\alpha_{12}$ measuring the (twice lagged) total indirect effect of the peers' $G$ can indeed be meaningfully compared to the similarly interpretable $TIL=SP\times DL=\rho_2\times\rho_3$ effect derived from the DGP that also occurs over two periods. When it comes to direct learning, both $\alpha_{13}$ and $\alpha_{23}$ are comparable to the true $DL=\rho_3$ from the DGP in \eqref{eq:omega_sim}. Tables \ref{tab:sim1s_alt1}--\ref{tab:sim1s_alt2} summarize these results. 

To examine the ability of alternative models to identify firm productivity, we first study if they can consistently estimate the production function coefficients (here $\beta_K$) because $\widehat{\omega}_{it}$ is a direct construct of these parameters: $\widehat{\omega}_{it}=y_{it}- \widehat{\beta}_Kk_{it}-\widehat{\beta}_Mm_{it}-\widehat{\eta}_{it}$.\footnote{The estimates of $\widehat{\beta}_M$ and $\widehat{\eta}_{it}$ are obtained from the material revenue share regression which does not depend on the Markovian assumption about $\omega_{it}$. Hence, they are exactly the same as those in our methodology.} The corresponding estimates of $\beta_K$ are reported in Table \ref{tab:sim1s_alt0} in Appendix \ref{sec:appendix_sim_nl}. These first-step results apply to both the ALT1 and ALT2 models and are obtained assuming that $\omega_{it}$ is an exogenous first-order Markov process. As expected, the estimation of production-function parameters (and hence, firm productivity) becomes biased with no tangible improvement following the growth in $n$ as soon as we deviate from the exogenous productivity process [scenarios (i) and (ii)]. In the latter case, biases originate from misspecification of the productivity proxy function that is missing relevant controls pertaining to productivity-modifying learning and/or spillovers.   

As seen in Tables \ref{tab:sim1s_alt1}--\ref{tab:sim1s_alt2}, the misestimation of productivity feeds into the second-step regressions. Across all experiments, both the ALT1 and ALT2 models exhibit non-vanishing biases in the estimation of spillovers. The same is also generally the case for estimation of within-firm direct learning, with the exception of the ALT2 estimator in the least probable scenarios when $G$ is an irrelevant uncorrelated covariate (i.e., when $DL=0$ \textit{and} $G$ evolves exogenously in population). Notably and perhaps more importantly, the alternative estimators fail at identifying (zero) cross-firm spillovers even when exogeneity of the Markov productivity process assumed in the first step is true [scenario (iii)]. This is because the second-step regressions remain misspecified due to their omission of the lagged productivity as customarily done in spillovers studies. Thus, if the first-step assumption of exogenous productivity in such analyses is indeed correct, the ``evidence'' of cross-firm spillovers uncovered in the second step is likely spurious and effectively driven by the missing \textit{auto}regressive dynamics in productivity \textit{within} the firm. This is not just a feature of specifications in \eqref{eq:alt1}--\eqref{eq:alt2}. In Appendix \ref{sec:appendix_sim_nl}, we consider their multiple variants drawn from the literature. Those results provide further evidence of the potential for spurious findings of spillovers using the popular two-step analysis procedure.

% ------------------------------------------------------------------------------------------
% ------------------------------------------------------------------------------------------

\section{Empirical Application}
\label{sec:fdiapplication}

We apply our methodology to study cross-firm spillovers with a particular focus on the productivity effects of inbound FDI via the domestic firms' learning of more advanced/efficient foreign knowledge. We proxy the firm's exposure to foreign knowledge using information on the share of foreign capital in its equity. This is a standard measure of foreign knowledge exposure in the literature. Thus, the foreign equity share $G_{it}\in[0,1]$ is our productivity modifier of interest.

Our objective is to study two potential channels---direct and indirect---of the productivity-boosting effects of inbound FDI. First, domestic firms may boost their productivity levels via ``importing'' better/new technology and learning more efficient management and marketing practices from abroad that they gain  \textit{direct} access to through foreign investors; these are direct technology transfers. A second mechanism by which domestic firms may \textit{indirectly} improve their productivity is by learning from other spatially proximate foreign-invested/owned firms in the industry and then adopting their superior practices already imported into the country. The latter channel is indirect and works through cross-firm peer effects. To model these indirect productivity effects of FDI, we need to explicitly recognize the potential for cross-sectional dependence in productivity which would permit FDI spillovers capable of influencing the domestic firms' productivity levels (and hence their output) beyond the immediate recipients. Our proposed model in Section \ref{sec:model} readily provides an empirical framework for this analysis. It allows identification of both the direct/internal ($DL$) and indirect/external ($TIL$) effects of inbound FDI in the presence of non-zero productivity spillovers ($SP$) among peer firms. In line with Remark \ref{rem:omegalag_vs_Glag}, we model ``FDI spillovers'' as operating through the firm's exposure to the average peer productivity, i.e., via ``productivity spillovers'' due to agglomeration externalities more broadly.

\paragraph{Data.} Our data come from the Chinese Industrial Enterprises Database survey conducted by China's National Bureau of Statistics. We focus on the electric machinery and equipment manufacturing industry, SIC 2-digit code 39. The rationale behind the choice of this industry is discussed in Appendix \ref{sec:indsutry}. Our sample period runs from 1998 to 2007, and the operational sample is an unbalanced panel of 23,720 firms with a total of 73,095 observations. In Appendix \ref{sec:appendix_data}, we provide the details of variable construction and describe the data.

We use postal codes to identify spatial neighbors included in each firm's peer group $\mathcal{L}(i,t)$. Peers are defined at the city level and at the level of the upper administrative division (provinces, autonomous regions, municipalities under the direct rule of government and special administrative regions) to allow for a broader geographical extent of spillovers while also respecting regulatory, administrative and cultural heterogeneity across regions. For the baseline results, the industrial scope of peer effects is defined at the level of the whole 2-digit industry. We consider a more granular definition of industrial similarity at the 4-digit level in robustness checks.

% ------------------------------------------------------------------------------------------
% ------------------------------------------------------------------------------------------

\subsection{Results}
\label{sec:results}

Owing to the nonparametric specification of the firm productivity process, we obtain observation-specific heterogeneous estimates of $SP_{it}$, $DL_{it}$ and $TIL_{it}$. We estimate the unknown $h(\cdot)$ via sieve methods using the popular second-degree polynomial series.\footnote{For instance, see \citet{deloeckeretal2016} or \citet{gnr2013}. We have also experimented with higher-order polynomials, and the results are very similar except somewhat noisier, as expected.} All estimations include time effects (the quadratic time trend yields qualitatively similar results). Also note that, because $\omega_{it}$ is the log-productivity, $SP_{it}$ is an elasticity measured in percents per unit percent of firm productivity, whereas both the $DL_{it}$ and $TIL_{it}$ are semi-elasticities  measured in percents per unit percentage \textit{point} change in the firm's foreign equity share. The measured learning effects on productivity are short-run and partial (i.e., for one given firm only). They do not capture mutual peer effects of an FDI injection across the network of firms, and neither do they account for dynamic effects over time. Obviously, owing to the persistence and cross-peer dependence of productivity, the cumulative implications of FDI for domestic firms' productivity in the long run equilibrium will be more sizable due to accumulation and diffusion over time and space.

Table \ref{tab:sp_dl_til__s} summarizes semiparametric point estimates of cross-firm productivity spillovers along with the direct and indirect effects of FDI on the productivity of domestic firms from our baseline specification,\footnote{The associated production function parameter estimates are $\widehat{\beta}_M=0.74$, $\widehat{\beta}_K=0.05$ and $\widehat{\beta}_L=0.12$ with the implied scale elasticity of $0.91$. These are in line with the Cobb-Douglas estimates for Chinese manufacturing reported in the literature \citep[e.g., see][]{brandetal2017} and suggest that the industry exhibits the decreasing returns to scale.} in which each firm's peer group is restricted to the same province and the industrial scope of spillovers is defined at the level of the entire 2-digit industry. All reported estimates are accompanied by the two-tailed 95\% bootstrap intervals. In addition, we formally test for significantly \textit{positive} productivity effects at each observation using the \textit{one}-sided 95\% bootstrap lower bounds. Throughout, we use accelerated bias-corrected bootstrap percentile confidence intervals (see Appendix \ref{sec:inference}). The share of firms for which the estimates statistically exceed zero are reported in the last column of Table \ref{tab:sp_dl_til__s}. In Appendix \ref{sec:appendix_emp}, we also summarize these productivity effect estimates graphically via empirical distributions across firms.

The estimated median $DL$ effect of own FDI is 0.14, whereby an increase of the foreign share in the median firm's equity by 10 percentage points boosts its productivity next year by 1.4\%. Expectedly, the $TIL$ effect of peers' FDI is smaller in magnitude\textemdash 0.04 at the median\textemdash so a 10 percentage point increase in the peer group average of the foreign equity share boosts the firm's productivity by only 0.4\%. Overall, at least 87\% of firms enjoy significant productivity-boosting effects of inbound FDI, both directly and indirectly. 

The non-zero external/indirect learning effect of FDI is facilitated by the presence of substantial and positive cross-firm productivity spillovers in the industry, with the median spillover elasticity $SP$ estimated at 0.33 along with the corresponding interquartile range of 0.18--0.45. Thus, a 10\% improvement in the average productivity of the firm's peers is estimated to increase the median firm's own productivity by about 3.3\%. These productivity spillovers are significantly positive for roughly 84\% of firms in the industry. We examine their geographic distribution in Appendix \ref{sec:appendix_emp}.

\paragraph{Heterogeneity and Nonlinearity.} Even within a given industry, firms are highly heterogeneous across many dimensions including their productivity and the extent of their exposure to foreign investors, both direct and through their peers. These characteristics can influence the effect size of spillovers and learning. Conveniently, our model readily facilitates testing of that.

Recall that we estimate the productivity effects of interest via $\widehat{SP}_{it}=\partial \widehat{h}_{it}(\cdot)/\partial \sum_{j(\ne i)}s_{ij,t-1}{\omega}_{j,t-1}$ and 
$\widehat{DL}_{it}=\partial \widehat{h}_{it}(\cdot)/\partial G_{i,t-1}$, where we estimate $h\left(\cdot\right)$ using the second-order polynomial sieve approximation. Thus, by analytical derivation, the estimated $\widehat{SP}_{it}$ and $\widehat{DL}_{it}$ are the linear functions of the ``determinants'' of firm productivity $(\omega_{i,t-1},G_{i,t-1},\sum_{j(\ne i)}s_{ij,t-1}{\omega}_{j,t-1})'$. Table \ref{tab:sp_dl_til__s_hetero} reports the estimates of sieve coefficients on these three variables in the $SP$ and $DL$ functions. 

Consider the spillovers first. The coefficient on the firm's own productivity is negative, indicating that the spillover effects decline in magnitude as firms themselves become more productive. Thus, less productive manufacturers have a greater potential to benefit from positive peer effects. Also consistent with economic intuition, the effect size of spillovers increases with the average productivity of peers: there is more to learn from highly productive neighbors in the industry. We also find a negative relationship between the firm's foreign equity share and the effect size of spillovers. This suggests that the domestic firms  experiencing larger productivity improvements via indirect learning from their foreign-invested peers\textemdash thanks to positive productivity spillovers\textemdash are those with limited direct access to foreign knowledge through their own investors (i.e., low-foreign-equity-share firms). 

In the case of FDI effects on productivity, results in the far right column of Table \ref{tab:sp_dl_til__s_hetero} suggest that the direct learning effects diminish as the firm's productivity rises, implying that the more productive firms have less absorptive capacity to learn. The foreign share in a firm's equity negatively affects the learning effect size, which basically indicates the diminishing productivity returns to receiving FDI.  Lastly, there is evidence that the higher the average of peer productivity, the lesser the productivity boosts from FDI. Thus, positive spillovers from highly productive neighbors essentially diminish the importance of direct FDI effects.

\paragraph{Robustness Analysis.} We first assess robustness of our empirical findings of significantly positive productivity spillovers and learning effects of FDI to the following modeling choices: (i) the inclusion of peer group effects to control for unobservable ``correlated effects;'' (ii) the composition of a reference peer group $\mathcal{L}(i,t)$; and (iii) the peer-weighting scheme $\{ s_{ijt}\}$.

As discussed in Section \ref{sec:model}, to structurally identify productivity spillovers $SP$, we rule out unobservable ``correlated effects'' at the peer group level. However, we can replace this no-group-effects assumption with a much milder assumption allowing for network unobservables but having them be time-invariant. In this case, we can control for the potential network confounders using group-level fixed effects \citep[see][]{grahamhahn2005,bramoulleetal2009}. We consider group effects across both the spatial and industrial dimensions. Specifically, we re-estimate our baseline specification by adding fixed effects at the level of the entire peer group as well as more granular subgroups.\footnote{These peer group effects are included in the second-stage estimation that models the productivity process.} The corresponding results are summarized in columns F1--F4 of Table \ref{tab:sp_dl_til__s_robust} (see table notes for the details on group fixed effects). While predictably there is no  dramatic change in the $DL$ estimates of within-firm learning, the median effect size of cross-firm spillovers $SP$ increases notably when we rely solely on the within-group variation over time to estimate the productivity peer effects.\footnote{Larger magnitudes of $TIL$ are a direct result of the increased $SP$ estimates.} The latter is especially true when the correlated group effects are defined narrowly at the 4-digit sub-industry level. In this case, the median spillover effect is 0.61 (against the baseline estimate of 0.33) and the effect is statistically positive for almost all firms (99\%). The increase in the effect size of spillovers is indicative of the substantial between-group heterogeneity in (peer) firm productivity, which is consistent with the well-documented differential in productivity levels across regions in China. Thus, when omitting group-specific effects, the measure of the strength of peer dependence across firms gets ``diluted'' in the baseline model due to the variation across groups.\footnote{This is the case even if the strength of within-group spillovers is the same for all groups.}

In columns P1--P3 of Table \ref{tab:sp_dl_til__s_robust}, we estimate productivity spillovers under three alternative definitions of who the firm's relevant peers are. Each one presumes a much smaller reference group than the baseline. Namely, we consider narrowing the scope of local spillovers to the level of city and/or the 4-digit sub-industry. The direct effect of the firm's own FDI expectedly continues to stay largely unchanged, but the estimates of productivity spillovers diminish in size significantly. The latter indirectly corroborates the rationale of our baseline specification in that the agglomeration effects have broad geographical and industrial scopes. By restricting the extent of spillovers to the local city and/or the firm's sub-industry only, we also restrict the reach of cross-firm externalities in productivity. Intuitively, when restricting the firm's learning opportunities to a narrower group of neighbors, a 10\% improvement in the average peer productivity is estimated to help boost the firm's own productivity by only about 0.6--0.9\% at the median. In contrast, if the relevant peer reference group is actually larger in scope, the same 10\% improvement across all peers (as in the baseline specification) implies a bigger industry-wide aggregate effect and, consequently, a larger estimated spillover effect on the firm of 3.3\%.

In Table \ref{tab:sp_dl_til__s_robust}, we also consider an alternative way of weighting peers, whereby bigger neighbors get assigned larger relative weights (column W1). The spillover effects only modestly decline in size. Overall, the cross-model variation in the spillover estimates we observe in Table \ref{tab:sp_dl_til__s_robust} is unsurprising and, in fact, expected because each model treats peer interactions a bit differently and/or utilizes different variation in data to identify productivity spillovers. Having said that, the $SP$ point estimates across all models are highly positively correlated, with the rank correlation coefficient being 0.81 on average. Consistently across all specifications, we continue to find that the overwhelming majority of the electric machinery manufacturers in China enjoy positive and significant productivity spillovers, in general, and FDI spillovers, specifically.

Appendix \ref{sec:appendix_emp} contains additional robustness checks, including to the potential violation of the weak exogeneity of the \textit{lagged} foreign equity share. Controlling for potential endogeneity of the FDI exposure, we continue to find strong evidence in support of significantly positive productivity spillovers for 80\% of firms or more, with our findings remaining qualitatively unchanged. 
In the same appendix, we also explore potential heterogeneity in the external productivity spillovers from the peers \textit{conditional} on their FDI status, which gives rise to \textit{bi}dimensional spillovers. We find evidence of heterogeneity in the strength of spillovers from wholly-domestic versus foreign-invested peers but, overall, our main findings stay the same: productivity spillovers are positive and significant for most firms. For the details, see Appendix \ref{sec:appendix_emp}.

% ------------------------------------------------------------------------------------------
% ------------------------------------------------------------------------------------------

\section{Conclusion}
\label{sec:conclusion}

This paper develops a novel methodology for the proxy variable structural identification of (latent) firm productivity in the presence of learning and cross-firm spillovers which allows a unified one-step analysis of the knowledge-transfer effects \textit{between} peer firms. Our framework is fundamentally different from the popular empirical approach traditionally implemented in two steps, whereby one first recovers firm productivity using the available standard proxy variable estimators and then tests for spillovers in the second step by  regressing these productivity estimates on various peer-group averages capturing firms' exposure to potential spillovers. Contrary to such an approach, our methodology is ``internally consistent'' in that it does not postulate contradictory assumptions. In building our model, we explicitly accommodate cross-sectional dependence in firm productivity induced by spillovers. We also show that estimating the firm production function or productivity using traditional proxy methods while ignoring the spillover-induced cross-sectional dependence, as customarily done in the literature, likely leads to misspecification and endogeneity-generating omitted variable bias. Because our methodology can be easily adapted to admit various spillover origins such as spatial agglomeration, R\&D, FDI, exporting, etc., it is fit to investigate cross-firm productivity spillovers in many contexts.

% ------------------------------------------------------------------------------------------
% ------------------------------------------------------------------------------------------
% ------------------------------------------------------------------------------------------
% ------------------------------------------------------------------------------------------
% ------------------------------------------------------------------------------------------

{%\doublespacing 
\onehalfspacing \small
\setlength{\bibsep}{0pt} \bibliography{Productivitybib} }

% ------------------------------------------------------------------------------------------
% ------------------------------------------------------------------------------------------
% ------------------------------------------------------------------------------------------
% ------------------------------------------------------------------------------------------
% ------------------------------------------------------------------------------------------

\clearpage
% -----------------------------------------------------
\begin{table}[p]
	\caption{Simulation Results for Our Estimation Methodology}\label{tab:sim1s}
	\centering
	\small
	\begin{tabular}{l r| rrr | rrr | rrr }
		\toprule[1pt]
		& True & \multicolumn{3}{c}{$n=100$} & \multicolumn{3}{c}{$n=200$} & \multicolumn{3}{c}{$n=400$} \\
		& Value & Mean & RMSE & MAE & Mean & RMSE & MAE & Mean & RMSE & MAE \\
		\midrule
		& \multicolumn{10}{c}{\sc DGP with $G$ Evolving Exogenously}  \\[2pt]
		\multicolumn{11}{l}{\bf Scenario (i): $DL\ne0$ and $SP\ne 0$ } \\[2pt]
		$\beta_K$ 		& 0.25 	& 0.249 & 0.055 & 0.043 & 0.250 & 0.038 & 0.030 & 0.252 & 0.027 & 0.021 \\
		$AR$ 			& 0.55	& 0.546 & 0.080 & 0.065 & 0.547 & 0.054 & 0.044 & 0.549 & 0.038 & 0.031 \\
		$DL$ 			& 0.50	& 0.500 & 0.164 & 0.134 & 0.502 & 0.113 & 0.091 & 0.501 & 0.079 & 0.064 \\ 
		$SP$ 			& 0.40	& 0.399 & 0.144 & 0.116 & 0.400 & 0.102 & 0.082 & 0.399 & 0.070 & 0.057 \\ 
		$TIL$ 			& 0.20	& 0.199 & 0.121 & 0.097 & 0.201 & 0.085 & 0.069 & 0.199 & 0.059 & 0.047 \\[2pt]
		   
		\multicolumn{11}{l}{\bf Scenario (ii): $DL=0$ and $SP\ne 0$ } \\[2pt]
		$\beta_K$ 		& 0.25	& 0.249 & 0.055 & 0.042 & 0.250 & 0.037 & 0.029 & 0.252 & 0.027 & 0.021 \\
		$AR$ 			& 0.55	& 0.546 & 0.085 & 0.070 & 0.546 & 0.059 & 0.048 & 0.549 & 0.042 & 0.034 \\
		$DL$ 			& 0		& --0.002 & 0.154 & 0.124 & 0.001 & 0.106 & 0.087 & 0.000 & 0.075 & 0.060 \\
		$SP$ 			& 0.40	& 0.395 & 0.172 & 0.140 & 0.402 & 0.124 & 0.102 & 0.400 & 0.085 & 0.069 \\
		$TIL$ 			& 0		& 0.004 & 0.067 & 0.049 & 0.002 & 0.045 & 0.034 & 0.000 & 0.030 & 0.024 \\[2pt]
		   
		\multicolumn{11}{l}{\bf Scenario (iii): $DL=0$ and $SP=0$ } \\[2pt]
		$\beta_K$ 		& 0.25	& 0.248 & 0.051 & 0.040 & 0.250 & 0.035 & 0.028 & 0.251 & 0.025 & 0.020 \\
		$AR$ 			& 0.55	& 0.546 & 0.083 & 0.067 & 0.546 & 0.058 & 0.047 & 0.549 & 0.040 & 0.033 \\
		$DL$ 			& 0		& 0.003 & 0.155 & 0.125 & 0.003 & 0.106 & 0.086 & 0.001 & 0.075 & 0.061 \\
		$SP$ 			& 0		& --0.031 & 0.157 & 0.131 & --0.009 & 0.108 & 0.091 & --0.004 & 0.078 & 0.065 \\
		$TIL$ 			& 0		& --0.004 & 0.028 & 0.018 & --0.002 & 0.013 & 0.008 & --0.001 & 0.006 & 0.004 \\
		\midrule
		& \multicolumn{10}{c}{\sc DGP with $G$ Following an $\omega$-Controlled Process}  \\[2pt]
		\multicolumn{11}{l}{\bf Scenario (i): $DL\ne0$ and $SP\ne 0$ } \\[2pt]
		$\beta_K$ 		& 0.25	& 0.249 & 0.060 & 0.047 & 0.251 & 0.040 & 0.032 & 0.252 & 0.029 & 0.023 \\
		$AR$ 			& 0.55	& 0.544 & 0.110 & 0.089 & 0.546 & 0.073 & 0.060 & 0.548 & 0.053 & 0.043 \\
		$DL$ 			& 0.50	& 0.503 & 0.148 & 0.120 & 0.502 & 0.099 & 0.081 & 0.502 & 0.070 & 0.057 \\
		$SP$ 			& 0.40	& 0.404 & 0.058 & 0.047 & 0.403 & 0.041 & 0.033 & 0.401 & 0.028 & 0.023 \\ 
		$TIL$ 			& 0.20	& 0.201 & 0.069 & 0.056 & 0.201 & 0.047 & 0.038 & 0.201 & 0.032 & 0.026 \\[2pt]
		   
		\multicolumn{11}{l}{\bf Scenario (ii): $DL=0$ and $SP\ne 0$ } \\[2pt]
		$\beta_K$ 		& 0.25	& 0.251 & 0.058 & 0.044 & 0.250 & 0.037 & 0.029 & 0.252 & 0.027 & 0.021 \\
		$AR$ 			& 0.55	& 0.544 & 0.105 & 0.087 & 0.545 & 0.070 & 0.058 & 0.548 & 0.051 & 0.042 \\
		$DL$ 			& 0		& 0.004 & 0.134 & 0.108 & 0.003 & 0.092 & 0.076 & 0.002 & 0.067 & 0.055 \\
		$SP$ 			& 0.40	& 0.387 & 0.224 & 0.183 & 0.400 & 0.158 & 0.130 & 0.399 & 0.116 & 0.094 \\
		$TIL$ 			& 0		& --0.007 & 0.060 & 0.043 & --0.003 & 0.038 & 0.028 & --0.001 & 0.027 & 0.021 \\[2pt]
		   
		\multicolumn{11}{l}{\bf Scenario (iii): $DL=0$ and $SP=0$ } \\[2pt]
		$\beta_K$ 		& 0.25	& 0.248 & 0.052 & 0.041 & 0.250 & 0.035 & 0.028 & 0.251 & 0.025 & 0.019 \\
		$AR$ 			& 0.55	& 0.546 & 0.101 & 0.083 & 0.546 & 0.069 & 0.056 & 0.548 & 0.050 & 0.040 \\
		$DL$ 			& 0		& --0.001 & 0.134 & 0.109 & 0.000 & 0.092 & 0.075 & 0.001 & 0.068 & 0.054 \\
		$SP$ 			& 0		& --0.031 & 0.155 & 0.128 & --0.008 & 0.105 & 0.088 & --0.004 & 0.075 & 0.063 \\
		$TIL\qquad$		& 0		& 0.000 & 0.022 & 0.015 & 0.000 & 0.011 & 0.007 & 0.000 & 0.005 & 0.003 \\
		\midrule
		\multicolumn{11}{p{15cm}}{\footnotesize {\sc Notes:} Owing to linearity of the productivity process in \eqref{eq:omega_sim}, the true values of $AR$, $DL$, $SP$ and $TIL$ are the fixed coefficients for all $i$ and $t$, and $TIL=SP\times DL$ is derived indirectly. Throughout, $T=10$. } \\
		\bottomrule[1pt] 
	\end{tabular}
\end{table}
% -----------------------------------------------------

\clearpage
% -----------------------------------------------------
\begin{table}[t]
	\caption{Simulation Results for the Two-Step Alternative Estimator of Spillovers ALT1}\label{tab:sim1s_alt1}
	\centering
	\small
	\begin{tabular}{l r| rr | rr | rr }
		\toprule[1pt]
		& True & \multicolumn{2}{c}{$n=100$} & \multicolumn{2}{c}{$n=200$} & \multicolumn{2}{c}{$n=400$} \\
		& Value & Mean & RMSE & Mean & RMSE & Mean & RMSE  \\
		\midrule
		& \multicolumn{7}{c}{\sc DGP with $G$ Evolving Exogenously}  \\[2pt]
		\multicolumn{8}{l}{\bf Scenario (i): $DL\ne0$ and $SP\ne 0$ } \\[2pt]
		$DL$ 			& 0.50	& 0.167 & 0.355 & 0.171 & 0.340 & 0.173 & 0.333 \\
		$TIL$ 			& 0.20	& --0.577 & 0.793 & --0.578 & 0.787 & --0.577 & 0.781 \\[2pt]
		   
		\multicolumn{8}{l}{\bf Scenario (ii): $DL=0$ and $SP\ne 0$ } \\[2pt]
		$DL$ 			& 0		& --0.202 & 0.225 & --0.200 & 0.211 & --0.200 & 0.206 \\
		$TIL$ 			& 0		& --0.348 & 0.372 & --0.345 & 0.361 & --0.352 & 0.357 \\[2pt]
		   
		\multicolumn{8}{l}{\bf Scenario (iii): $DL=0$ and $SP=0$ } \\[2pt]
		$DL$ 			& 0		& 0.461 & 0.470 & 0.464 & 0.468 & 0.465 & 0.467 \\
		$TIL$ 			& 0		& 0.165 & 0.204 & 0.170 & 0.191 & 0.169 & 0.180 \\
		\midrule
		& \multicolumn{7}{c}{\sc DGP with $G$ Following an $\omega$-Controlled Process}  \\[2pt]
		\multicolumn{8}{l}{\bf Scenario (i): $DL\ne0$ and $SP\ne 0$ } \\[2pt]
		$DL$ 			& 0.50	& 1.651 & 1.152 & 1.650 & 1.151 & 1.652 & 1.152 \\
		$TIL$ 			& 0.20	& 0.582 & 0.387 & 0.585 & 0.387 & 0.586 & 0.387 \\[2pt]
		   
		\multicolumn{8}{l}{\bf Scenario (ii): $DL=0$ and $SP\ne 0$ } \\[2pt]
		$DL$ 			& 0		& 0.416 & 0.420 & 0.420 & 0.422 & 0.424 & 0.425 \\
		$TIL$ 			& 0		& --0.112 & 0.119 & --0.113 & 0.117 & --0.115 & 0.117 \\[2pt]
		   
		\multicolumn{8}{l}{\bf Scenario (iii): $DL=0$ and $SP=0$ } \\[2pt]
		$DL$ 			& 0		& 0.616 & 0.619 & 0.617 & 0.619 & 0.621 & 0.621 \\
		$TIL\qquad$		& 0		& --0.362 & 0.365 & --0.361 & 0.362 & --0.362 & 0.363 \\
		\midrule
		\multicolumn{8}{p{11.2cm}}{\footnotesize {\sc Notes:} Reported are the results from the second-step regression in \eqref{eq:alt1} estimated with $\widehat{\omega}_{it}$ obtained in the first step using the standard proxy estimator under the assumption of exogenous Markov productivity process. $DL$ and $TIL$ are respectively measured by $\alpha_{13}$ and $\alpha_{12}$, with the latter capturing spillovers. Throughout, $T=10$. } \\
		\bottomrule[1pt] 
	\end{tabular}
\end{table}
% -----------------------------------------------------

\clearpage
% -----------------------------------------------------
\begin{table}[t]
	\caption{Simulation Results for the Two-Step Alternative Estimator of Spillovers ALT2}\label{tab:sim1s_alt2}
	\centering
	\small
	\begin{tabular}{l r| rr | rr | rr }
		\toprule[1pt]
		& True & \multicolumn{2}{c}{$n=100$} & \multicolumn{2}{c}{$n=200$} & \multicolumn{2}{c}{$n=400$} \\
		& Value & Mean & RMSE & Mean & RMSE & Mean & RMSE  \\
		\midrule
		& \multicolumn{7}{c}{\sc DGP with $G$ Evolving Exogenously}  \\[2pt]
		\multicolumn{8}{l}{\bf Scenario (i): $DL\ne0$ and $SP\ne 0$ } \\[2pt]
		$DL$ 			& 0.50	& 0.800 & 0.313 & 0.814 & 0.320 & 0.817 & 0.320 \\
		$SP$ 			& 0.40	& 1.074 & 0.676 & 1.083 & 0.684 & 1.090 & 0.690 \\[2pt]
		   
		\multicolumn{8}{l}{\bf Scenario (ii): $DL=0$ and $SP\ne 0$ } \\[2pt]
		$DL$ 			& 0		& --0.036 & 0.093 & --0.015 & 0.063 & --0.008 & 0.044 \\
		$SP$ 			& 0.40	& 0.751 & 0.518 & 0.874 & 0.502 & 0.924 & 0.527 \\[2pt]
		   
		\multicolumn{8}{l}{\bf Scenario (iii): $DL=0$ and $SP=0$ } \\[2pt]
		$DL$ 			& 0		& 0.005 & 0.084 & 0.004 & 0.059 & 0.003 & 0.042 \\
		$SP$ 			& 0		& 0.539 & 0.541 & 0.545 & 0.546 & 0.548 & 0.548 \\
		\midrule
		& \multicolumn{7}{c}{\sc DGP with $G$ Following an $\omega$-Controlled Process}  \\[2pt]
		\multicolumn{8}{l}{\bf Scenario (i): $DL\ne0$ and $SP\ne 0$ } \\[2pt]
		$DL$ 			& 0.50	& 1.149 & 0.650 & 1.152 & 0.652 & 1.156 & 0.656 \\
		$SP$ 			& 0.40	& 0.580 & 0.182 & 0.579 & 0.180 & 0.577 & 0.178 \\[2pt]
		   
		\multicolumn{8}{l}{\bf Scenario (ii): $DL=0$ and $SP\ne 0$ } \\[2pt]
		$DL$ 			& 0		& 0.369 & 0.376 & 0.376 & 0.379 & 0.375 & 0.377 \\
		$SP$ 			& 0.40	& --0.708 & 1.304 & --0.635 & 1.121 & --0.583 & 1.020 \\[2pt]
		   
		\multicolumn{8}{l}{\bf Scenario (iii): $DL=0$ and $SP=0$ } \\[2pt]
		$DL$ 			& 0		& 0.406 & 0.406 & 0.405 & 0.406 & 0.407 & 0.407 \\
		$SP\qquad$ 			& 0		& 0.390 & 0.393 & 0.395 & 0.396 & 0.397 & 0.397 \\
		\midrule
		\multicolumn{8}{p{11.2cm}}{\footnotesize {\sc Notes:} Reported are the results from the second-step regression in \eqref{eq:alt2} estimated with $\widehat{\omega}_{it}$ obtained in the first step using the standard proxy estimator under the assumption of exogenous Markov productivity process. $DL$ and $SP$ are respectively measured by $\alpha_{23}$ and $SP=\alpha_{22}$. Throughout, $T=10$. } \\
		\bottomrule[1pt] 
	\end{tabular}
\end{table}
% -----------------------------------------------------

\clearpage
% ---------------------------------------
\begin{table}[t]
	\centering
	\caption{Estimates of the Productivity Effects}
	\label{tab:sp_dl_til__s}
	\small
	\makebox[\linewidth]{
		\begin{tabular}{l ccc|c}
			\toprule
			& \multicolumn{3}{c}{\textit{Point Estimates}} & {\textit{Statistically $>0$}} \\   
			Estimand & 1st Qu. & Median & 3rd Qu. & (\% Obs.)  \\
			\midrule
			$SP$ 	& 0.182 & 0.327 & 0.452 & 83.84 \\
			& (0.108, 0.240) & (0.211, 0.399) & (0.302, 0.543) &  \\
			$DL$ 	& 0.096 & 0.138 & 0.168 & 89.13 \\
			& (0.076, 0.118) & (0.118, 0.164) & (0.146, 0.197) &  \\	
			$TIL$ 	& 0.022 & 0.037 & 0.053 & 86.71 \\
			& (0.011, 0.032) & (0.022, 0.050) & (0.034, 0.068) &  \\
			\midrule
			\multicolumn{5}{p{12cm}}{\footnotesize {\sc Notes:} Reported are the results based on our baseline specification of the productivity process in \eqref{eq:proddist_a_exp}. The left panel summarizes point estimates of $SP_{it}$, $DL_{it}$ and $TIL_{it}$ with the corresponding two-sided 95\% bootstrap percentile confidence intervals in parentheses. The last column reports the share of observations for which the point estimates are statistically positive at the 5\% significance level using one-sided bootstrap confidence intervals.} \\
			\bottomrule[1pt]
		\end{tabular}
	}
\end{table}
% ---------------------------------------

\clearpage
% -----------------------------------
\begin{table}[t]
	\caption{Heterogeneity and Nonlinearity in the Productivity Effects}
	\label{tab:sp_dl_til__s_hetero}
	\centering	\small
	\begin{tabular}{lcc}
		\toprule[1pt]
		& $SP$ & $DL$ \\
		\midrule
		$\omega_{i,t-1} $ 		 & --0.735 & --0.164 \\
		& (--0.890, --0.550)	& (--0.203, --0.126) \\
		$G_{i,t-1}$ 			 &  --0.198 & --0.155 \\
		& (--0.403, --0.103)	& (--0.184, --0.127) \\
		$\sum_j s_{ij,t-1}\omega_{j,t-1}$ 	& 1.249 & --0.198 \\
		& (0.423, 1.527) 	& (--0.403, --0.103)\\
		\midrule
		\multicolumn{3}{p{8.4cm}}{\footnotesize {\sc Notes:} Reported are the parameter estimates for the $SP$ and $DL$ functions derived from the polynomial approximation of the conditional mean of $\omega_{it}$ in the productivity process formulation in \eqref{eq:proddist_a_exp}. Two-sided 95\% bootstrap percentile confidence intervals in parentheses. These correspond to our baseline specification. }\\
		\bottomrule[1pt]
	\end{tabular}
\end{table} 
% -----------------------------------

\clearpage
% ---------------------------------------
\begin{sidewaystable}[p]
	\centering
	\caption{Estimates of the Productivity Effects: Robustness Analysis}
	\label{tab:sp_dl_til__s_robust}
	\small
	\makebox[\linewidth]{
		\begin{tabular}{l cccc | c | ccc}
			\toprule
			Estimand & F1 & F2 & F3 & F4 & W1 & P1 & P2 & P3 \\   
			\midrule
			&\multicolumn{8}{c}{\textbf{---Median Estimates---}} \\[2pt]
			$SP$	& 0.499 & 0.410 & 0.607 & 0.605 & 0.222 & 0.076 & 0.085 & 0.064 \\
			& (0.386, 0.535) & (0.313, 0.428) & (0.540, 0.614) & (0.524, 0.647) & (0.186, 0.254) & (0.048, 0.100) & (0.049, 0.122) & (0.051, 0.077) \\
			$DL$	& 0.167 & 0.176 & 0.171 & 0.177 & 0.144 & 0.146 & 0.145 & 0.149 \\
			& (0.145, 0.194) & (0.155, 0.205) & (0.149, 0.198) & (0.156, 0.205) & (0.123, 0.171) & (0.123, 0.172) & (0.121, 0.171) & (0.128, 0.174) \\
			$TIL$	& 0.071 & 0.060 & 0.089 & 0.092 & 0.009 & 0.009 & 0.010 & 0.008 \\
			& (0.050, 0.079) & (0.046, 0.067) & (0.074, 0.100) & (0.076, 0.106) & (0.006, 0.014) & (0.006, 0.013) & (0.006, 0.016) & (0.006, 0.011) \\
			\midrule
			&\multicolumn{8}{c}{\textbf{---Statistically $>0$ (\% Obs.)---}} \\[2pt]
			$SP$	& 95.98 & 94.19 & 98.78 & 99.15 & 77.96 & 92.25 & 89.69 & 88.89 \\
			$DL$	& 90.24 & 90.39 & 90.34 & 90.43 & 89.11 & 89.43 & 89.69 & 89.57 \\
			$TIL$	& 96.31 & 94.72 & 99.07 & 99.31 & 72.46 & 93.11 & 89.34 & 94.97 \\
			\midrule
			\multicolumn{9}{p{22cm}}{\footnotesize {\sc Notes:} Reported are the results for the productivity process formulation in \eqref{eq:proddist_a_exp}. The two-sided 95\% bootstrap percentile confidence intervals for the median point estimates are in parentheses. Statistical positiveness is at the 5\% significance level, using one-sided bootstrap percentile confidence intervals. In the baseline specification: (i) each firm's peers are restricted to the firms located in the same province and the industrial scope of spillovers defined at the level of the entire 2-digit industry, (ii) the technical change is flexibly controlled for using a series of year effects. The other model specifications are based on the baseline except for the following alternate features. 
				F1: includes location fixed effects at the level of peer groups (here, province); 
				F2: includes location fixed effects at the level of peer subgroups (here, city); 
				F3: includes location-industry fixed effects at the level of peer subgroups (here, province and 4-digit sub-industry); 
				F4: includes spatial and industry fixed effects at the level of peer subgroups (here, city and 4-digit sub-industry); 
				W1: peers are size-weighted, i.e., $s_{ijt}= {L_{jt}\mathbbm{1}\big\{(j,t)\in \mathcal{L}(i,t)\big\} }\big/{\sum_k L_{kt} \mathbbm{1}\big\{(k,t)\in \mathcal{L}(i,t)\big\} }$; 
				P1: peers in $\mathcal{L}(i,t)$ are restricted to the 4-digit sub-industry; 
				P2: peers in $\mathcal{L}(i,t)$ are restricted to the same city; 	
				P3: peers in $\mathcal{L}(i,t)$ are restricted to the 4-digit sub-industry \textit{and} the same city. } \\
			\bottomrule[1pt]
		\end{tabular}
	}
\end{sidewaystable}
% ---------------------------------------

% ------------------------------------------------------------------------------------------
% ------------------------------------------------------------------------------------------
% ------------------------------------------------------------------------------------------
% ------------------------------------------------------------------------------------------
% ------------------------------------------------------------------------------------------
% ------------------------------------------------------------------------------------------
% ------------------------------------------------------------------------------------------
% ------------------------------------------------------------------------------------------
% ------------------------------------------------------------------------------------------
% ------------------------------------------------------------------------------------------
% ------------------------------------------------------------------------------------------
% ------------------------------------------------------------------------------------------
% ------------------------------------------------------------------------------------------
% ------------------------------------------------------------------------------------------
% ------------------------------------------------------------------------------------------

\clearpage
\thispagestyle{empty}\setcounter{page}{1}\setcounter{footnote}{0}

\

\vspace{5cm}
\begin{center}	
	{\it \LARGE Online Appendix to} \\ 
	{ \huge  On the Estimation of Cross-Firm Productivity \\ Spillovers with an Application to FDI} \\
	
	\vspace{1cm}
	
	{\sc \large Emir Malikov$^1$ $\qquad $ Shunan Zhao$^2$} \\
	\vspace{0.5cm}
	{\small $^1$University of Nevada, Las Vegas  \\ $^2$Oakland University}	
\end{center} 

\normalsize 

% ------------------------------------------------------------------------------------------
% ------------------------------------------------------------------------------------------

\clearpage
\appendix

\setlength{\abovedisplayskip}{3pt}
\setlength{\belowdisplayskip}{3pt} 

\setcounter{figure}{0} \renewcommand\thetable{\thesection.\arabic{table}}
\setcounter{table}{0} \renewcommand\thefigure{\thesection.\arabic{figure}}

\section{Relation to the Augmented Production Function Approach}
\label{sec:appendix_augment}

A two-step framework, which we seek to improve upon in this paper, is not universal across empirical studies of productivity spillovers. The exceptions are predominantly from the literature on R\&D-borne productivity spillovers, where some studies instead adopt a singe-step methodology centered on the estimation of the \citet{griliches1979}-style ``augmented production function'' which, besides the conventional inputs, also explicitly admits the firm's own and external knowledge capital stock. Seemingly, such a model readily provides estimates of the ``contextual'' spillover effects of R\&D on firm production in one step. However, this framework is rather unique to the studies of spillovers in R\&D, because this productivity-enhancing activity is the most input-accumulation-like in that it is an investment into the knowledge capital. Augmenting the firm's production function to include the FDI/exports/imports variables and their respective spillover pool measures is however not as conceptually unambiguous. Specifically, this is problematic on at least two fronts. First, the spillover effects on firm productivity in such a setup is essentially assumed to be deterministic, whereby the impact on productivity is improbably the same for all firms without a possibility of the varying degree of success (say, due to random luck or misfortune). Second, including internal and external measures of productivity modifiers directly into the production function effectively implies substitutability of the firm's inputs with not only its own productivity-enhancing activities such FDI or exporting but also\textemdash and perhaps more eyebrow-raising\textemdash with those of its peers. This remark equally applies to the case of R\&D spillovers\footnote{More recently, the literature has been gravitating towards embedding the firm's R\&D behavior into the productivity process and taking it out of the production function itself; e.g., see \citet{dj2013,dj2018}.} and is along the lines of \citeauthor{deloecker2013}'s (2013) critique in the context of estimating the learning-by-exporting effects on firm productivity. Not least importantly, identification of productivity spillovers in prior studies (including those via R\&D) may also be seriously hindered by the well-known econometric problems with standard proxy-based or (dynamic panel) fixed-effects production function estimators.\footnote{See \citet{gm1998}, \citet{acf2015} and \citet{gnr2013} for the discussion of various production-function estimators and identification challenges associated with them.} This further highlights the practical usefulness of our proposed methodology.

% ------------------------------------------------------------------------------------------
% ------------------------------------------------------------------------------------------

\section{China's Electric Machinery Manufacturing}
\label{sec:indsutry}

Our empirical analysis focuses on China's electric machinery and equipment manufacturing industry which includes manufacturing of generators and motors, power transmission and distribution equipment, wires and cables, batteries, household electric and non-electric appliances, lighting appliances, etc. We select this industry because it is histrionically one of the country's most fundamental manufacturing sectors. The development of this industry has been closely related to the growth of GDP and the ever-expanding demand of electricity. By its very nature, the industry has thus been crucial for promoting the overall industrialization in China. Besides that, electric machinery and equipment is also China's most exported product \citep{euroeximbank}, and the industry amounts to over a quarter of global sales \citep{deloitte}. It is also one of the manufacturing industries that receive most of FDI. For instance, in 2005 alone (near the end of our sample period) the machinery and equipment industry in China attracted \$4 billion in foreign investment, which was about 10\% of FDI inflows to Chinese manufacturing that year \citep{comm}.

Foreign-invested firms are the dominant players in this industry. Arguably, this is mainly due to China's lack of domestic innovation capabilities, excessive failure rates of R\&D and, consequently, high dependence on new technologies imported from abroad, particularly during the first decade following renewed privatization efforts in the late 1990s (the period of our analysis). For example, according to the Xiamen Bureau of Statistics, in Xiamen (a large and important port-city on the East coast) just 48 large foreign-invested firms owned 82\% of fixed assets and produced 79\% of the output value in the local electric machinery industry in 2005. 

More generally, the Chinese electric machinery and equipment manufacturing industry is characterized by a high degree of spatial clustering and industrial agglomeration [mainly on the coast; see Figure \ref{fig:numberoffirms_map}(a) in Appendix \ref{sec:appendix_data}] typical for such technology- and skill-intensive industries. Along with the government's emphasis on innovations and new technologies as a means for sustainable development of this industry, this makes it an interesting application for studying productivity effects of inbound FDI and the associated spillovers across firms.

% ------------------------------------------------------------------------------------------
% ------------------------------------------------------------------------------------------

\section{Translog Production Function}
\label{sec:appendix_translog}

Our methodology can adapt more flexible specifications of the firm's production function. The log-quadratic translog specification provides a natural extension of the log-linear Cobb-Douglas form that we have assumed in \eqref{eq:prodfn}. The former is more flexible and implies input and scale elasticities that vary both over time and across firms thereby being more robust to firm heterogeneity. For instance, see \citet{deloeckerwarzynski2012} and \citet{deloeckeretal2016} for recent applications of the translog production functions in the structural proxy estimation.

Let the firm's stochastic production function takes the following form \textit{in logs}:
\begin{align}\label{eq:prodfn_gross_logs*}
y_{it} =&\ \beta_0 + \beta_Kk_{it}+\tfrac{1}{2}\beta_{KK}k_{it}^2+ \beta_Ll_{it}+\tfrac{1}{2}\beta_{LL}l_{it}^2 +\beta_Mm_{it}+\tfrac{1}{2}\beta_{MM}m_{it}^2\ + \notag \\ 
&\ \beta_{KL}k_{it}l_{it}+\beta_{KM}k_{it}m_{it}+\beta_{LM}l_{it}m_{it}+ \omega_{it} + \eta_{it} \notag \\
\equiv&\ T(k_{it},l_{it},m_{it})+ \omega_{it} + \eta_{it} ,
\end{align}
where $T(k_{it},l_{it},m_{it})$ is a shorthand for the translog expansion of inputs. All the remaining assumptions about the market environment, productivity processes, timing of production decisions and learning, etc.~stay unchanged.

The firm's static optimization problem with respect to materials now is
\begin{align}\label{eq:profitmax*}
\max_{M_{it}}\ P_{t}^Y \exp\{ T(k_{it},l_{it},m_{it}) \} \exp\{\omega_{it}\}\theta  - P_{t}^M M_{it} ,
\end{align}
with the corresponding first-order condition given by
\begin{equation}\label{eq:profitmax_foc*}
P_{t}^Y \exp\{ T(k_{it},l_{it},m_{it}) \} 
\frac{\beta_M+\beta_{MM}m_{it}+\beta_{KM}k_{it}+\beta_{LM}l_{it}}{M_{it}}\exp\{\omega_{it}\}\theta = P_{t}^M .
\end{equation}

Dividing \eqref{eq:profitmax_foc*} by the translog production function expressed in levels and then taking logs of both sides, we obtain the following material share equation:
\begin{equation}\label{eq:fst*}
\ln V_{it} = \ln (\left[\beta_M+\beta_{MM}m_{it}+\beta_{KM}k_{it}+\beta_{LM}l_{it} \right]\theta) - \eta_{it} ,
\end{equation}
where $\beta_M+\beta_{MM}m_{it}+\beta_{KM}k_{it}+\beta_{LM}l_{it}$ is the material elasticity function. Analogous to the discussion in Section \ref{sec:identification}, the above share equation identifies the material-related production-function parameters $(\beta_M,\beta_{MM},\beta_{KM},\beta_{LM})'$ as well as the mean of exponentiated shocks $\theta=\mathbb{E}[ \exp\{\eta_{it}\}]$ based on the mean-orthogonality condition $\mathbb{E}[\eta_{it} |\ \mathcal{I}_{it}] = \mathbb{E}[\eta_{it}] = 0$. These parameters are to be estimated in the first stage via nonlinear least squares on \eqref{eq:fst*}.

Having identified the production function in the dimension of its endogenous static input $m_{it}$, we focus on the remaining production-function parameters as well as the nonparametric evolution process for $\omega_{it}$. With the already identified $y_{it}^*\equiv y_{it} - \beta_Mm_{it}-\tfrac{1}{2}\beta_{MM}m_{it}^2-\beta_{KM}k_{it}m_{it}-\beta_{LM}l_{it}m_{it}$ and using the Markovian process for productivity, we now have the analogue of \eqref{eq:prodfn_gross_logs_3}:
\begin{equation}\label{eq:prodfn_gross_logs_3*}
y_{it}^* = \beta_Kk_{it}+\tfrac{1}{2}\beta_{KK}k_{it}^2+ \beta_Ll_{it}+\tfrac{1}{2}\beta_{LL}l_{it}^2 + \beta_{KL}k_{it}l_{it} + h\left(\omega_{i,t-1}, G_{i,t-1}, \sum_{j(\ne i)}s_{ij,t-1}\omega_{j,t-1}\right) + \zeta_{it} + \eta_{it} 
\end{equation}
that contains no endogenous variables on the right-hand side. Next, proxying for $\omega_{i,t-1}$ and $\omega_{j,t-1}$ via the inverted material function derived from \eqref{eq:profitmax_foc*}, we obtain
\begin{align}\label{eq:sst*}
	y_{it}^* =&\ \beta_Kk_{it}+\tfrac{1}{2}\beta_{KK}k_{it}^2+ \beta_Ll_{it}+\tfrac{1}{2}\beta_{LL}l_{it}^2 + \beta_{KL}k_{it}l_{it}\ +   \\
	&\ h\Bigg(\omega_{i,t-1}^*\left(\beta_K,\beta_L,\beta_{KK},\beta_{LL},\beta_{KL}\right), G_{i,t-1},  \sum_{j(\ne i)}s_{ij,t-1}\omega_{j,t-1}^*\left(\beta_K,\beta_L,\beta_{KK},\beta_{LL},\beta_{KL}\right)\Bigg) +  \zeta_{it} + \eta_{it},  \notag 
\end{align}
where the productivity proxy function is given by
\begin{align}\label{eq:omega_starr*}
	\omega_{it}^*\left(\beta_K,\beta_L,\beta_{KK},\beta_{LL},\beta_{KL}\right) = \varkappa_{it}^*- \beta_Kk_{it}-\tfrac{1}{2}\beta_{KK}k_{it}^2- \beta_Ll_{it}-\tfrac{1}{2}\beta_{LL}l_{it}^2 - \beta_{KL}k_{it}l_{it}\quad \forall i,t, 
\end{align}
with 
\begin{align*}
	\varkappa_{it}^*=&\ \ln(P_t^M/P_t^Y)-\ln (\left[\beta_M+\beta_{MM}m_{it}+\beta_{KM}k_{it}+\beta_{LM}l_{it} \right]\theta)\ - \\ 
	&\ (1-\beta_M)m_{it}-\tfrac{1}{2}\beta_{MM}m_{it}^2-\beta_{KM}k_{it}m_{it}-\beta_{LM}l_{it}m_{it}
\end{align*}
being a function of the parameters that have already been identified in the first stage.

A semiparametric model in \eqref{eq:sst*} is then identified based on the same moment restriction as in \eqref{eq:sst_ident}, with all right-hand-side covariates being weakly exogenous and thus self-instrumenting. Approximating the unknown $h(\cdot)$ via linear sieves, \eqref{eq:sst*} is to be estimated in the second stage via semiparametric nonlinear least-squares. The remaining aspects closely follow the estimation procedure outlined in Section \ref{sec:estimation}.

% ------------------------------------------------------------------------------------------
% ------------------------------------------------------------------------------------------

\section{Asymmetric Productivity Spillovers}
\label{sec:appendix_asym}

Our baseline peer weighing scheme in \eqref{eq:weight1} treats cross-firm spillovers symmetrically in that all members of a peer group affect each other's productivity. That is, each $i$th firm's productivity is influenced by the average productivity of all its peers: those that are more \textit{and} those that are less productive than the firm $i$ itself. Given that we have no prior beliefs about the directionality of productivity spillovers in China's electric machinery manufacturing that we study in our empirical application, we opt for a symmetric specification. But should one choose to regulate the direction of productivity spillovers by restricting them to occur \textit{from} more productive \textit{to} less productive firms, our framework can be modified to accommodate that too. 
	
The latter case however implies a somewhat different conceptualization of cross-firm dependence in which firms are said to learn exclusively from (relative) productivity ``leaders.'' The identification of such \textit{a}symmetric spillovers, which are conditional on the firm's own productivity relative to that of its peers, generally requires additional structural/timing assumptions. 
	
To model productivity spillovers between firms asymmetrically, we can redefine peer weights $\{s_{ijt}\}$ as follows:
\begin{equation}\label{eq:weight1*}
		s_{ijt}^*= \frac{\mathbbm{1}\big\{(j,t)\in \mathcal{L}(i,t)\ \text{and}\ \omega_{j,t-1}>\omega_{i,t-1}\big\} }
		{\sum_{k(\ne i)=1}^{n} \mathbbm{1}\big\{(k,t)\in \mathcal{L}(i,t)\ \text{and}\ \omega_{j,t-1}>\omega_{i,t-1}\big\} },
\end{equation}
so that only the neighbors who are more productive than the firm $i$ are identified as its peers for external cross-firm learning. Note that, in the above, the relevant peers at time $t$ are selected based on their relative productivity superiority in the \textit{previous} period $t-1$. Without this, we would not be able to separate the cross-firm spillover effect from the firm's own autoregressive effect, conflating the two. The latter becomes obvious when we substitute \eqref{eq:weight1*} into the Markov productivity process \eqref{eq:proddist_a_exp} that describes the evolution of firm $i$'s productivity over time:
\begin{equation}\label{eq:proddist_a_exp*}
	\omega_{it} = \mathbb{E}\Bigg[\omega_{it} \Bigg|\ \omega_{i,t-1}, G_{i,t-1}, \sum_{j(\ne i)}\underbrace{\frac{\mathbbm{1}\big\{(j,t-1)\in \mathcal{L}(i,t-1)\ \text{and}\ \omega_{j,t-2}>\omega_{i,t-2}\big\} }
		{\sum_{k(\ne i)=1}^{n} \mathbbm{1}\big\{(k,t-1)\in \mathcal{L}(i,t-1)\ \text{and}\ \omega_{j,t-2}>\omega_{i,t-2}\big\} }}_{s_{ij,t-1}^*}\omega_{j,t-1}\Bigg] + \zeta_{it}.
\end{equation}
	
By making the asymmetry in external learning be a function of the \textit{twice}-lagged pair-wise productivity differentials between the firm and its peers, we avoid the appearance of $\omega_{i,t-1}$ in two places thereby allowing us to partial out the cross-firm spillovers from the \textit{auto}regressive persistence in productivity.
Thus, to separably identify asymmetric spillovers in productivity, in addition to assuming that (both the internal and external) learning occurs with a delay, one also requires an assumption that the firm takes an additional period to identify more productive peers. However, we do not need this additional timing assumption in our baseline analysis (with symmetric interactions).

% ------------------------------------------------------------------------------------------
% ------------------------------------------------------------------------------------------

\section{Additional Modeling Considerations}
\label{sec:appx_add_ident}

\paragraph{Contextual Effects.} Because we assume delayed cross-firm peer interactions, as noted by \citet{manski1993}, the \textit{dynamic} nature of productivity model \eqref{eq:proddist_a_exp} can potentially provide an additional avenue to circumvent the unidentification problems and separate different types of peer effects, should one be interested in also modeling the ``contextual effects'' on productivity via $\sum_{j(\ne i)}s_{ij,t-1}G_{j,t-1}$. In such a setup, per the results in \citet{bramoulleetal2009}, the separable identification of ``endogenous''  and ``contextual'' effects can also be achieved by relying on variation in the size of peer reference groups, so long as the firm $i$ is excluded when computing group means, as is in our case. Alternatively, (co)variance-based quadratic moment conditions may be used to aid identification \citep{kp1999,lee2007,kp2020}.

\paragraph{Contemporaneous Effects.} Depending on the particular source of learning, it may sometimes be possible to reasonably relax the timing assumption that the learning effect of $G_{it}$ on firm productivity be with a delay. Take, for example, the firm's export status in the context of ``learning by exporting.'' Consistent with much theoretical and empirical work in international trade, the decision to start exporting is usually associated with large sunk entry costs, which would impede firms from adjusting their export status \textit{immediately} after experiencing an improvement in their productivity. Analogous arguments can be made about costliness of swift geographic relocations. If so, it may be feasible to replace weak exogeneity of lagged $G_{i,t-1}$ and  $\{s_{ij,t-1}\}$ with a stronger assumption of weak exogeneity of $G_{it}$ and  $\{s_{ijt}\}$. The productivity process in \eqref{eq:proddist_a_exp} can then be modified as follows: $\omega_{it} = \mathbb{E}\left[\omega_{it} |\ \omega_{i,t-1}, G_{it}, \sum_{j(\ne i)}s_{ijt}\omega_{j,t-1}\right] + \zeta_{it}$, where the implied mean-orthogonality of $\zeta_{it}$ and $(G_{it}, \sum_{j(\ne i)}s_{ijt}\omega_{j,t-1})'$ is effectively paramount to assuming that, due to adjustment costs, both the $G_{it}$ and firm location in period $t$ are determined in period $t-1$ based on $\omega_{i,t-1}$ just like the dynamic inputs are.

% ------------------------------------------------------------------------------------------
% ------------------------------------------------------------------------------------------

\section{Inference}
\label{sec:inference}

\paragraph{Asymptotic Inference.} Let the moment vector in \eqref{eq:msys} be concisely written as $\mathbb{E}[\boldsymbol{\rho}({\Theta})]=\mathbf{0}$, where $\Theta$ is a collection of both the finite-dimensional coefficients $\left(\beta_M,\beta_K,\beta_L,\theta\right)'$ and nonparametric sieve ``parameters'' $\boldsymbol{\gamma}$. Given the just-identification of the model and so long as we use linear sieves for $\mathcal{A}_{L_{n}}\left(\cdot\right)$ such as polynomial or B-spline series, we can make use of the numerical equivalence \citep[see][]{hahnetal2018} between the consistent estimator of the asymptotic variance of \textit{semiparametric} ``parameters'' $\widehat{{\Theta}}$ and a consistent estimator of the asymptotic variance derived for these parameter estimators as if the estimated model were of a \textit{parametric} form specified in \eqref{eq:fst_est} and \eqref{eq:sst_est}. Thus, in practice, one can use the variance formula for a parametric two-step estimator to consistently estimate the variance of a semiparametric sieve estimator.\footnote{Note that this equivalence applies to finite samples only because, asymptotically, the number of sieve ``parameters'' will diverge to infinity with the sample size whereas the number of parameters in a parametric specification will stay a finite constant. Furthermore, the numerical equivalence holds more generally for fully \textit{non}parametric two-step sieve estimators. In our case, the estimator is \textit{semi}parametric, with the first step implemented using the known parametric form. Since ours is a special case of the nonparametric setup studied by \citet{hahnetal2018}, their results continue to apply.} The asymptotic variance for such a parametric two-step estimator can in turn be derived following \citeauthor{newey1984}'s (1984) suggestion by making use of the optimal GMM covariance formula: $\mathbb{V}ar\big[\widehat{{\Theta}}\big] = \big[\mathbb{E}\frac{\partial \boldsymbol{\rho}({\Theta})}{\partial {\Theta}' }\big]^{-1} \mathbb{E} [\boldsymbol{\rho}({\Theta})\boldsymbol{\rho}({\Theta})' ]
\big[\mathbb{E}\frac{\partial \boldsymbol{\rho}({\Theta})}{\partial {\Theta} }\big]^{-1}$. This streamlines asymptotic inference. 

\paragraph{Bias-Corrected Bootstrap Inference.} However, because asymptotic inference for semi- and nonparametric estimators is well-known to perform unreliably due to finite-sample biases as well as the first-order asymptotic theory's poor ability to approximate the distribution of estimators in finite samples \citep{horowitz2001}, for hypothesis testing, we therefore rely on \citeauthor{efron1987}'s (1987) accelerated bias-corrected bootstrap percentile confidence intervals, which are second-order accurate and provide means not only to correct for the estimator's finite-sample bias but also to account for higher-order moments (particularly, skewness) in the sampling distribution.

We approximate sampling distributions of the estimator via wild residual block bootstrap that takes into account a panel structure of the data, with both stages resampled jointly owing to a sequential nature of our estimation procedure. More specifically, when constructing wild bootstrap residuals, we work with the \textit{joint} distribution of firm-specific time series of $\{\widehat{\eta}_{it}\}$ and $\{\widehat{\zeta}_{it}\}$, with the auxiliary random variable drawn from the \citet{mammen1993} two-point distribution independently over $i$. Note that this independence over $i$ is consistent with our model's assumption about random productivity shocks. We set the number of bootstrap replications to $B=400$. Having first obtained bootstrap parameter estimates $\{(\widehat{\beta}_K^b,\widehat{\beta}_L^b,\widehat{\beta}_M^b)';\ b=1,\dots,B\}$ and $\{\widehat{\boldsymbol{\gamma}}^b;\ b=1,\dots,B\}$, we then obtain bootstrap values for our main estimands of interest: $\widehat{DL}_{it}^b$,  $\widehat{SP}_{it}^b$ and $\widehat{TIL}_{it}^b$ for $b=1,\dots,B$ (at each observation). Next, we use the accelerated bias-correction method to make inference about $DL$, $SP$ and $TIL$.

To make matters concrete, let the (observation-specific) estimand of focus be denoted by $\widehat{E}$.
We use the empirical distribution of $B$ bootstrap estimates $\big\{\widehat{E}^1,\dots, \widehat{E}^B\big\}$ to estimate $(1-a)\times100$\% confidence bounds for $\widehat{E}$ as intervals between the $[a_1\times100]$th and $[a_2\times100]$th percentiles of its bootstrap distribution with
\begin{align}
	a_1 = \Phi\left(\widehat{\phi}_0+\frac{\widehat{\phi}_0+\phi_{a/2}} {1-\widehat{c} \left(\widehat{\phi}_0+\phi_{a/2}\right)}\right)\quad\text{and}\quad
	a_2 = \Phi\left(\widehat{\phi}_0+\frac{\widehat{\phi}_0+\phi_{(1-a/2)}} {1-\widehat{c}\left(\widehat{\phi}_0+\phi_{(1-a/2)}\right)}\right),
\end{align}
where $\Phi(\cdot)$ is the standard normal cdf, $\phi_{\alpha}$ is the $(\alpha\times100)$th percentile of the standard normal distribution, 
\begin{align}
	\widehat{\phi}_0 = \Phi^{-1}\left(\#\big\{\widehat{E}^b<\widehat{E}\big\}/B\right)
\end{align}
is a bias-correction factor, and $\widehat{c}$ is an acceleration parameter which, following the literature, is estimated via jackknife as follows \citep[e.g., see][]{shaotu1995}:
\begin{align}
	\widehat{c} = \frac{\sum_{j=1}^J\left( \sum_{s=1}^J\widehat{E}^s - \widehat{E}^j\right)^3} {6\Big[ \sum_{j=1}^J\left( \sum_{s=1}^J\widehat{E}^s - \widehat{E}^j \right)^2 \Big]^{3/2}},
\end{align}
where $\widehat{E}^j$ is the $j(=1,\dots,J)$th jackknife estimate of $E$.\footnote{We have tried different versions of jackknife with similar results. We settle on a delete-$50T$ jackknife (i.e., leave-$50$-cross-sections-out) which respects the panel structure of our data while yielding a reasonable number of subsamples the estimation on which is not computationally prohibitive.}

Note that both the acceleration and bias-correction factors are different for each estimator, denoted here generically by $\widehat{E}$. That is, the bias-correction procedure is not only estimand-specific but may also be observation-specific as is in our case. Also, the estimated confidence intervals may not contain the original estimates if the finite-sample bias is large. 

% ------------------------------------------------------------------------------------------
% ------------------------------------------------------------------------------------------

\section{Additional Simulation Results}
\label{sec:appendix_sim_nl}

% -----------------------------------------------------
\begin{table}[p]
	\caption{Simulation Results for Our Estimator under the Nonlinear Productivity Process}\label{tab:sim1g}
	\centering
	\small
	\begin{tabular}{l r| rrr | rrr | rrr }
		\toprule[1pt]
		& Mean True & \multicolumn{3}{c}{$n=100$} & \multicolumn{3}{c}{$n=200$} & \multicolumn{3}{c}{$n=400$} \\
		& Value & Mean & RMSE & MAE & Mean & RMSE & MAE & Mean & RMSE & MAE \\
		\midrule
		& \multicolumn{10}{c}{\sc DGP with $G$ Evolving Exogenously}  \\[2pt]
		\multicolumn{11}{l}{\bf Scenario (i): $DL\ne0$ and $SP\ne 0$ } \\[2pt]
		$\beta_K$ 		& 0.250	& 0.237 & 0.077 & 0.059 & 0.247 & 0.046 & 0.035 & 0.251 & 0.031 & 0.024  \\
		$AR$ 			& 0.596	& 0.593 & 0.081 & 0.066 & 0.592 & 0.054 & 0.044 & 0.594 & 0.038 & 0.031 \\
		$DL$ 			& 0.412	& 0.416 & 0.157 & 0.128 & 0.418 & 0.106 & 0.087 & 0.414 & 0.075 & 0.061 \\
		$SP$ 			& 0.301	& 0.274 & 0.313 & 0.257 & 0.303 & 0.200 & 0.166 & 0.302 & 0.138 & 0.115 \\
		$TIL$ 			& 0.124	& 0.102 & 0.144 & 0.112 & 0.119 & 0.085 & 0.069 & 0.120 & 0.059 & 0.048 \\[2pt]
		   
		\multicolumn{11}{l}{\bf Scenario (ii): $DL=0$ and $SP\ne 0$ } \\[2pt]
		$\beta_K$ 		& 0.250	& 0.233 & 0.082 & 0.063 & 0.247 & 0.049 & 0.037 & 0.251 & 0.033 & 0.025 \\
		$AR$ 			& 0.609	& 0.606 & 0.081 & 0.067 & 0.606 & 0.055 & 0.045 & 0.608 & 0.039 & 0.032 \\
		$DL$ 			& 0		& 0.003 & 0.158 & 0.127 & 0.003 & 0.107 & 0.087 & 0.001 & 0.076 & 0.061 \\
		$SP$ 			& 0.274	& 0.088 & 0.301 & 0.246 & 0.222 & 0.198 & 0.162 & 0.257 & 0.130 & 0.107 \\
		$TIL$ 			& 0		& --0.011 & 0.075 & 0.049 & --0.005 & 0.038 & 0.026 & --0.002 & 0.023 & 0.017 \\[2pt]
		   
		\multicolumn{11}{l}{\bf Scenario (iii): $DL=0$ and $SP=0$ } \\[2pt]
		$\beta_K$ 		& 0.250	& 0.246 & 0.064 & 0.051 & 0.249 & 0.044 & 0.035 & 0.251 & 0.032 & 0.025 \\
		$AR$ 			& 0.672	& 0.623 & 0.078 & 0.064 & 0.623 & 0.054 & 0.044 & 0.626 & 0.038 & 0.031 \\
		$DL$ 			& 0		& 0.003 & 0.155 & 0.126 & 0.003 & 0.106 & 0.087 & 0.001 & 0.075 & 0.061 \\
		$SP$ 			& 0		& --0.034 & 0.152 & 0.126 & --0.010 & 0.106 & 0.086 & --0.004 & 0.073 & 0.060 \\
		$TIL$ 			& 0		& --0.004 & 0.027 & 0.017 & --0.002 & 0.012 & 0.008 & --0.001 & 0.006 & 0.004 \\
		\midrule
		& \multicolumn{10}{c}{\sc DGP with $G$ Following an $\omega$-Controlled Process}  \\[2pt]
		\multicolumn{11}{l}{\bf Scenario (i): $DL\ne0$ and $SP\ne 0$ } \\[2pt]
		$\beta_K$ 		& 0.250 & 0.249 & 0.028 & 0.022 & 0.250 & 0.020 & 0.015 & 0.251 & 0.014 & 0.011 \\
		$AR$ 			& 0.338	& 0.328 & 0.078 & 0.060 & 0.334 & 0.054 & 0.043 & 0.338 & 0.038 & 0.030 \\
		$DL$ 			& 1.148	& 1.157 & 0.119 & 0.093 & 0.152 & 0.083 & 0.064 & 0.148 & 0.058 & 0.046 \\
		$SP$ 			& 0.695	& 0.697 & 0.036 & 0.029 & 0.697 & 0.025 & 0.021 & 0.695 & 0.017 & 0.014 \\
		$TIL$ 			& 0.798	& 0.808 & 0.294 & 0.166 & 0.803 & 0.207 & 0.119 & 0.798 & 0.146 & 0.082 \\[2pt]
		   
		\multicolumn{11}{l}{\bf Scenario (ii): $DL=0$ and $SP\ne 0$ } \\[2pt]
		$\beta_K$ 		& 0.250	& 0.233 & 0.086 & 0.065 & 0.245 & 0.053 & 0.039 & 0.251 & 0.033 & 0.025 \\
		$AR$ 			& 0.609	& 0.609	& 0.102 & 0.084 & 0.607 & 0.068 & 0.055 & 0.608 & 0.050 & 0.040 \\
		$DL$ 			& 0		& --0.006 & 0.135 & 0.109 & --0.002 & 0.092 & 0.075 & 0.000 & 0.066 & 0.054 \\
		$SP$ 			& 0.274	& 0.085 & 0.341 & 0.277 & 0.207 & 0.215 & 0.175 & 0.259 & 0.152 & 0.124 \\
		$TIL$ 			& 0		& 0.023 & 0.069 & 0.045 & 0.005 & 0.034 & 0.023 & 0.003 & 0.021 & 0.015 \\[2pt]
		   
		\multicolumn{11}{l}{\bf Scenario (iii): $DL=0$ and $SP=0$ } \\[2pt]
		$\beta_K$ 		& 0.250	& 0.246 & 0.068 & 0.053 & 0.249 & 0.044 & 0.035 & 0.251 & 0.031 & 0.025 \\
		$AR$ 			& 0.627	& 0.624 & 0.100 & 0.082 & 0.623 & 0.067 & 0.054 & 0.626 & 0.049 & 0.040 \\
		$DL$ 			& 0		& --0.001 & 0.134 & 0.108 & 0.001 & 0.092 & 0.075 & 0.001 & 0.066 &  0.054\\
		$SP$ 			& 0		& --0.035 & 0.147 & 0.122 & --0.009 & 0.101 & 0.084 & --0.004 & 0.071 & 0.060 \\
		$TIL\qquad$		& 0		& 0.002 & 0.021 & 0.014 & 0.000 & 0.010 & 0.007 & 0.000 & 0.005 & 0.003 \\
		\midrule
		\multicolumn{11}{p{15.9cm}}{\footnotesize {\sc Notes:} Owing to nonlinearity of the productivity process in \eqref{eq:omega_sim_nl}, $AR$, $DL$, $SP$ and $TIL$ are all observation-specific. With the sole exception for the fixed parameter $\beta_K=0.25$, the mean true values are the averages (across simulation repetitions) of the mean simulated values over $i$ and $t$. Throughout, $T=10$. } \\
		\bottomrule[1pt] 
	\end{tabular}
\end{table}
% -----------------------------------------------------

\paragraph{Nonlinear Productivity Process.} Table \ref{tab:sim1g} presents the results for our proposed estimator when the productivity DGP is nonlinear. Specifically, we consider the following nonlinear productivity process:
\begin{align}\label{eq:omega_sim_nl}
	\omega_{it}=&\ \rho_0+\rho_{11}\omega_{i,t-1}+\rho_{12}\omega_{i,t-1}^2
	+\rho_{21}\sum_{j(\ne i)}s_{ij,t-1}\omega_{j,t-1}+\rho_{22}\Bigg(\sum_{j(\ne i)}s_{ij,t-1}\omega_{j,t-1}\Bigg)^2\ + \notag \\
	&\ \rho_{31}G_{i,t-1}+\rho_{32}G_{i,t-1}^2+
	\varrho_{12}\omega_{i,t-1}\Bigg(\sum_{j(\ne i)}s_{ij,t-1}\omega_{j,t-1}\Bigg) + \varrho_{13}\omega_{i,t-1}G_{i,t-1}\ + \notag \\
	&\ \lambda_{23}G_{i,t-1}\Bigg(\sum_{j(\ne i)}s_{ij,t-1}\omega_{j,t-1}\Bigg) + 	\zeta_{it},
\end{align}
where $\rho_0=0.2$, $\rho_{11}=0.65$, $\rho_{12}=-0.015$, $\rho_{21}=0.18$, $\rho_{22}=0.025$, $\rho_{31}=0.37$, $\rho_{32}=0.12$, $\varrho_{12}=0.006$, $\varrho_{13}=-0.06$ and $\lambda_{23}=0.07$. The rest of the DGP is kept unchanged (see Section \ref{sec:simulations}).

Table \ref{tab:sim1g} essentially replicates Table \ref{tab:sim1s} using this new productivity process. The simulation results remain encouraging and show that our estimation methodology is consistent and recovers the true parameters well.

% -----------------------------------------------------
\begin{table}[t]
	\caption{Simulation Results for the Alternative Estimator of $\beta_K$}\label{tab:sim1s_alt0}
	\centering
	\small
	\begin{tabular}{l r| rr | rr | rr }
		\toprule[1pt]
		& True & \multicolumn{2}{c}{$n=100$} & \multicolumn{2}{c}{$n=200$} & \multicolumn{2}{c}{$n=400$} \\
		& Value & Mean & RMSE & Mean & RMSE & Mean & RMSE  \\
		\midrule
		& \multicolumn{7}{c}{\sc DGP with $G$ Evolving Exogenously}  \\[2pt]
		Scenario (i): $DL\ne0$ and $SP\ne 0$ 	& 0.25 	& 0.376 & 0.136 & 0.373 & 0.128 & 0.374 & 0.126 \\
		Scenario (ii): $DL=0$ and $SP\ne 0$ 	& 0.25	& 0.438 & 0.193 & 0.435 & 0.188 & 0.436 & 0.187 \\
		Scenario (iii): $DL=0$ and $SP=0$ 		& 0.25	& 0.251 & 0.040 & 0.250 & 0.029 & 0.251 & 0.020 \\
		\midrule
		& \multicolumn{7}{c}{\sc DGP with $G$ Following an $\omega$-Controlled Process}  \\[2pt]
		Scenario (i): $DL\ne0$ and $SP\ne 0$ & 0.25	& 0.118 & 0.152 & 0.113 & 0.148 & 0.109 & 0.146 \\
		\midrule 
		\multicolumn{8}{p{14.6cm}}{\footnotesize {\sc Notes:} Reported are the first-step results for $\widehat{\beta}_K$ from the alternative estimators which proxy for latent productivity under the assumption of exogenous Markov process for $\omega_{it}$. The results corresponding to scenarios (ii) and (iii) of the second DGP [bottom panel] are omitted because they are identical to those for the first DGP [top panel]. This is because not only does $G$ not enter the alternative estimator but it also does not affect the evolution of firm productivity by design ($DL=0$) in these two scenarios. Throughout, $T=10$. } \\
		\bottomrule[1pt] 
	\end{tabular}
\end{table}
% -----------------------------------------------------

\paragraph{First-Step Estimates of $\beta_K$ from the Two-Step Estimator.} To examine the ability of alternative models to identify firm productivity, we first study if these two-step estimators can consistently estimate the production function coefficients (here $\beta_K$) because $\widehat{\omega}_{it}$ is a direct construct of these parameters. The corresponding estimates of $\beta_K$ are reported in Table \ref{tab:sim1s_alt0}. These first-step results apply to both the ALT1 and ALT2 models and are obtained assuming that $\omega_{it}$ is an exogenous first-order Markov process.

% -----------------------------------------------------
\begin{table}[t]
	\caption{Simulation Results for Different Variants of the Two-Step Alternative Estimator}\label{tab:sim1s_alt3}
	\centering
	\small
	\makebox[\linewidth]{\begin{tabular}{l  rrrrrrr | rrrr }
		\toprule[1pt]
		& \multicolumn{7}{c}{Results for $\alpha_{12}$ ($TIL$) in eq.~\eqref{eq:alt1}} & 
		\multicolumn{4}{c}{Results for $\alpha_{22}$ ($SP$) in eq.~\eqref{eq:alt2}} \\ 
				 	& I & II & III & IV & V & VI & VII & VIII & IX & X & XI \\
		\midrule  
		\multicolumn{12}{l}{\bf Mean Estimate } \\ 
		$n=100$		& 1.154 & 0.558 & 1.057 & 0.620 & --0.359 & 0.378 & --0.362 & 0.990 & 0.591 & 0.540 & 0.390 \\
		$n=200$		& 1.157 & 0.556 & 1.064 & 0.624 & --0.355 & 0.386 & --0.361 & 0.995 & 0.597 & 0.546 & 0.395 \\
		$n=400$		& 1.167 & 0.560 & 1.073 & 0.628 & --0.356 & 0.387 & --0.362 & 0.998 & 0.599 & 0.548 & 0.397 \\[4pt]
		\multicolumn{12}{l}{\bf Root Mean Squared Error } \\ 
		$n=100$		& 1.171 & 0.596 & 1.072 & 0.647 & 0.359 & 0.416 & 0.365 & 0.987 & 0.595 & 0.541 & 0.393 \\
		$n=200$		& 1.166 & 0.575 & 1.072 & 0.638 & 0.356 & 0.405 & 0.362 & 0.995 & 0.599 & 0.547 & 0.396 \\
		$n=400$		& 1.170 & 0.570 & 1.077 & 0.635 & 0.357 & 0.398 & 0.363 & 0.998 & 0.600 & 0.549 & 0.397 \\[4pt]
		\multicolumn{12}{l}{\bf Rejection Frequency for $H_0: \text{\normalfont Spillover Parameter} =0$} \\ 
		$n=100$		& 1.00 & 0.97 & 1.00 & 0.99 & 1.00 & 0.90 & 1.00 & 1.00 & 0.99 & 1.00 & 1.00 \\
		$n=200$		& 1.00 & 1.00 & 1.00 & 1.00 & 1.00 & 0.98 & 1.00 & 1.00 & 1.00 & 1.00 & 1.00 \\
		$n=400$		& 1.00 & 1.00 & 1.00 & 1.00 & 1.00 & 1.00 & 1.00 & 1.00 & 1.00 & 1.00 & 1.00 \\
		\midrule
		\multicolumn{12}{l}{\bf Variables in the Second-Step Regression} \\ 
		Spillover Variable	& $\overline{G}_{it}$ & $\overline{G}_{it}$ 
		& $\overline{G}_{it-1}$ & $\overline{G}_{it-1}$
		& $\overline{G}_{it-2}$ & $\overline{G}_{it-2}$ 
		& $\overline{G}_{it-2}$ 
		& $\overline{\omega}_{it}$ & $\overline{\omega}_{it}$ & $\overline{\omega}_{it-1}$ & $\overline{\omega}_{it-1}$ \\
		DL Variable & -- & $G_{it}$ & -- & $G_{it-1}$ & -- & $G_{it-2}$ & $G_{it-1}$ 
		& -- & $G_{it}$ & -- & $G_{it-1}$ \\  
		\midrule
		\multicolumn{12}{p{17.2cm}}{\footnotesize {\sc Notes:} Reported are the results for ``spillovers'' (defined as either $SP$ or $TIL$) from different variants of the second-step regressions in \eqref{eq:alt1}--\eqref{eq:alt2} estimated with $\widehat{\omega}_{it}$ obtained in the first step using the standard proxy estimator under the assumption of exogenous Markov productivity. The data are simulated assuming linear productivity process under scenario (iii) with true $DL=0$ and $SP=0$. Thus, the true spillovers are zero. For the estimation of second-step regressions, the $G$ series is generated as an $\omega$-controlled process. The specifications containing contemporaneously endogenous regressors are estimated using first lags as instruments. $\overline{G}_{it}\equiv \sum_j s_{ijt}G_{jt}$ and $\overline{\omega}_{it}\equiv \sum_j s_{ijt}\omega_{jt}$. Throughout, $T=10$. } \\
		\bottomrule[1pt] 
	\end{tabular}}
\end{table}
% -----------------------------------------------------

\paragraph{Alternative Two-Step Estimators of Spillovers.}  Table \ref{tab:sim1s_alt3} reports the results for the ``spillovers'' estimator (defined as either $SP$ or $TIL$) from different variants of the second-step regressions in \eqref{eq:alt1}--\eqref{eq:alt2} estimated with $\widehat{\omega}_{it}$ obtained in the first step using the standard proxy estimator under the assumption of exogenous Markov productivity process. The data are simulated assuming a linear productivity process under scenario (iii) with the true $DL=0$ and $SP=0$. Thus, the first-step estimation of productivity is correctly specified and consistent. For the estimation of second-step regressions, the $G$ series is generated as an $\omega$-controlled process (b). The specifications containing contemporaneously endogenous regressors are estimated using their respective first lags as predetermined instruments. 

Examining the results in Table \ref{tab:sim1s_alt3}, we find that, across all specifications, the second-step estimator exhibits non-vanishing biases in the estimation of spillovers. All models spuriously fail at identifying \textit{zero} cross-firm spillovers. Here we also report the rejection frequencies (over simulation repetitions) for the asymptotic $z$-test of the null that the coefficient of a spillover variable in the model is zero at the 95\% confidence level. Had the second-step been correctly specified and consistent, we were to expect these rejection frequencies to be all around 0.05. Consistent with our expectations, the results in the table indicate drastic size distortions due to misspecification of the two-step approach. The results are qualitatively the same when we also control for firm and/or time fixed effects.

% ------------------------------------------------------------------------------------------
% ------------------------------------------------------------------------------------------

\section{Data}
\label{sec:appendix_data}

Our data are drawn from the Chinese Industrial Enterprises Database survey conducted by China's National Bureau of Statistics (NBS). This database covers all firms with sales above 5 million yuan (about 0.6 million in U.S. dollar) and includes most industries including mining, manufacturing and public utilities. We focus on the electric machinery and equipment manufacturing industry, SIC 2-digit code 39.

% ---------------------------------------
\begin{table}[t]
	\centering
	\caption{Data Summary Statistics}\label{tab:data}
	\small
	\makebox[\linewidth]{
		\begin{tabular}{lrrrr}
			\toprule
			Variable & Mean & 1st Qu. & Median  & 3rd Qu.\\
			\midrule
			$Y$     & 80,031 & 9,165  & 19,577 & 51,320 \\
			$K$     & 14,330 & 1,039  & 2,957  & 8,929 \\
			$L$     & 3,627  & 533   & 1,128  & 2,707 \\
			$M$     & 55,761 & 6,372  & 13,640 & 35,878 \\
			$G$     & 0.09 & 0 & 0 & 0 \\
			$\mathbbm{1}\{G>0\}\quad$ & 0.19 & & & \\
			\midrule
			\multicolumn{5}{p{7.8cm}}{\footnotesize {\sc Notes:} $Y$ -- gross output; $K$ -- capital stock; $L$ -- labor; $M$ -- materials; $G$ -- foreign equity share; $\mathbbm{1}\{G>0\}$ -- binary indicator signifying foreign-invested firms. $Y$, $K$, $L$ and $M$ are measured in thousands of RMB; $G$ is a unit-free proportion.} \\
			\bottomrule[1pt]
		\end{tabular}
	}
\end{table}
% ---------------------------------------

The production variables are as follows. The firm's capital stock ($K_{it}$) is the net fixed assets deflated by the price index of investment into fixed assets. Labor ($L_{it}$) is measured as the total wage bill plus benefits deflated by the GDP deflator. Materials ($M_{it}$) are the total intermediate inputs, including raw materials and other production-related inputs, deflated by the purchasing price index for industrial inputs. The output ($Y_{it}$) is defined as the gross industrial output value deflated by the producer price index. The price indices are obtained from NBS and the World Bank. The four variables are measured in thousands of real RMB. In addition, the foreign equity share ($G_{it}$) is a bounded proportion that lies between zero and one, by construction. 

% -----------------------------------
\begin{figure}[p]
	\centering
	\includegraphics[scale=0.5]{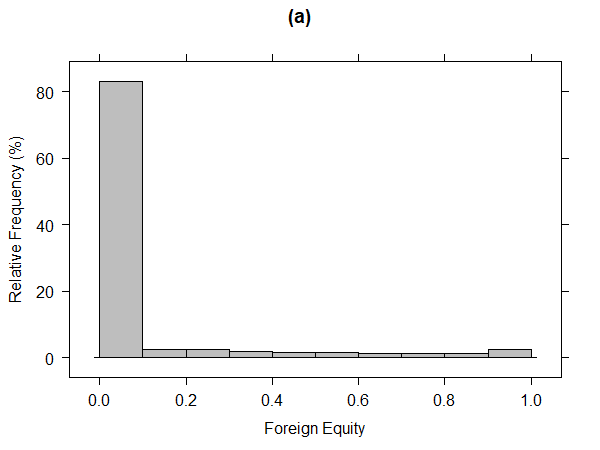}
	\includegraphics[scale=0.5]{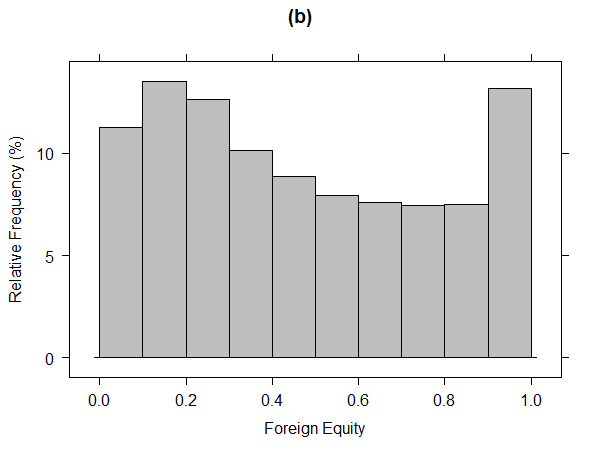}
	\caption{Empirical Distribution of the Foreign Equity Share:\\ (a) All Firms; (b) Foreign-Invested Firms Only}
	\label{fig:foreignequityshare}
\end{figure}
% -----------------------------------

% -----------------------------------
\begin{figure}[p]
	\centering
	\includegraphics[scale=0.5]{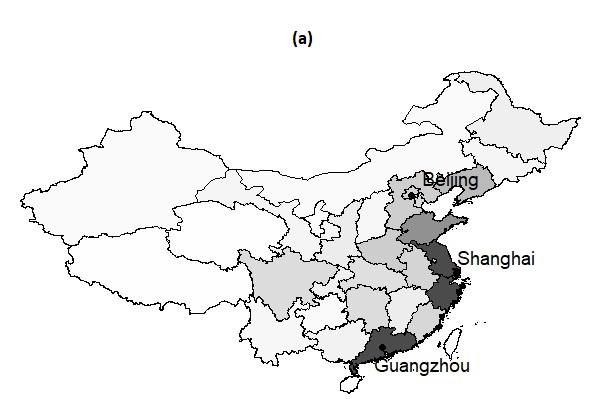}
	\includegraphics[scale=0.5]{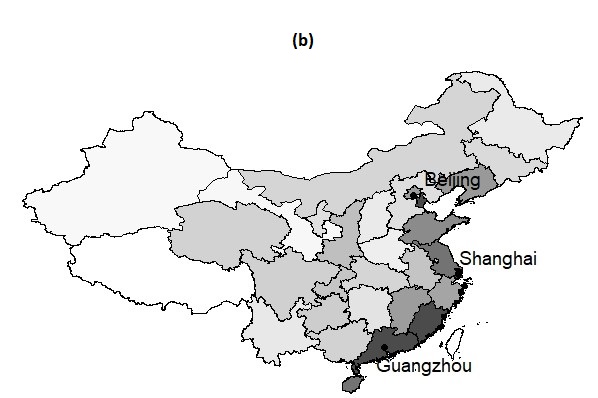}
	\caption{Geographic Distribution of Firms (a) and the Foreign Equity Share (b) \\ {\footnotesize {\sc Note:} Shown are the number of firms in the sample from each province (a) and the sample mean at the province level (b), with the darker areas corresponding to higher values.}}
	\label{fig:numberoffirms_map}
\end{figure}
% -----------------------------------

We exclude observations with missing values for these variables as well as a small number of likely erroneous observations with the foreign equity share values outside the unit interval. With the sample period running from 1998 to 2007, the operational sample is an unbalanced panel of 23,720 firms with a total of 73,095 observations. Table \ref{tab:data} reports summary statistics for these data.

Figure \ref{fig:foreignequityshare}(a) plots a histogram of $G_{it}$ across firms which expectedly has a zero mode because the manufacturing sector in China is dominated by wholly domestic firms. Figure \ref{fig:foreignequityshare}(b) on the right plots the distribution of $G_{it}|G_{it}>0$, i.e., for foreign-invested firms only. Overall, 81\% firms in our sample are wholly domestically-owned, 1\% are pure foreign multinationals, with the remaining 18\% being (partially) foreign-invested domestic firms. The map in Figure \ref{fig:numberoffirms_map}(b) shows the spatial distribution of the (average) foreign equity share across regions where, consistent with one's priors, we see the heightened concentration of FDI along the coast.

% ------------------------------------------------------------------------------------------
% ------------------------------------------------------------------------------------------

\section{Additional Empirical Results}
\label{sec:appendix_emp}

\begin{figure}[p]
	\centering
	\includegraphics[scale=0.5]{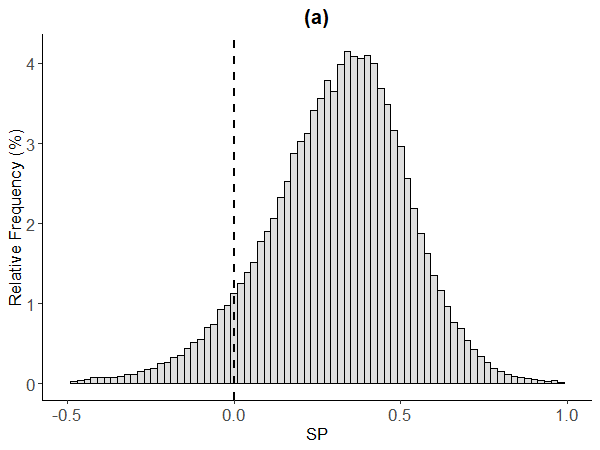}
	\includegraphics[scale=0.5]{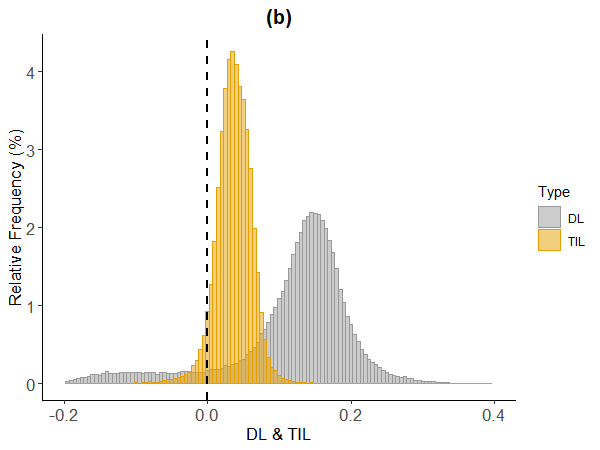}
	\caption{Distributions of the Productivity Effects: (a) $SP$, (b) $DL$ and $TIL$ \\ {\footnotesize {\sc Note:} Plotted are the point estimates based on our baseline specification of the productivity process in \eqref{eq:proddist_a_exp}.}}
	\label{fig:hist_sp_dl_til}
\end{figure}

\paragraph{Baseline Results.} Figure \ref{fig:hist_sp_dl_til} plots empirical histograms of the point estimates of productivity effects under the baseline specification. These estimates are the same as those summarized in Table \ref{tab:sp_dl_til__s}. Subfigure \ref{fig:hist_sp_dl_til}(a) shows the distribution of productivity spillover elasticities $SP$; the direct/internal and indirect/external learning effects of FDI ($DL$ and $TIL$, respectively) are presented in subfigure \ref{fig:hist_sp_dl_til}(b).

% ---------------------------------------
% ---------------------------------------

% ---------------------------------------
\begin{figure}[p]
	\centering
	\includegraphics[scale=0.55]{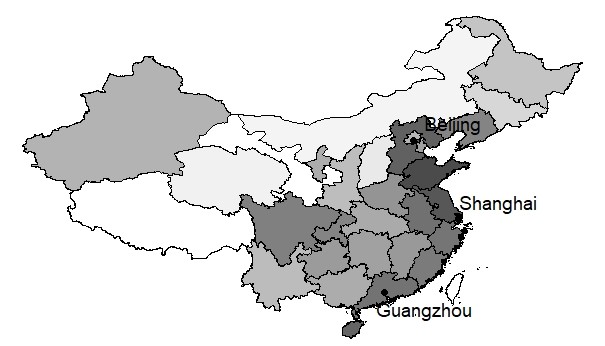}
	\caption{
		Spatial Distribution of the Productivity Spillover Effects\\
		{\footnotesize {\sc Note:} Plotted are the sample medians at the province level, with the darker areas corresponding to higher values.}	}
	\label{fig:map_sp}
\end{figure}
% ---------------------------------------

We also examine the geographic distribution of productivity spillovers in Figure \ref{fig:map_sp}. The map shows by-province median estimates of productivity spillovers. We observe that productivity spillovers are stronger in the highly industrialized, fast-growing provinces in the Southeast and near the coast. Interestingly, comparing Figure \ref{fig:map_sp} with the spatial distribution of the industry in Figure \ref{fig:numberoffirms_map}(a) in Appendix \ref{sec:appendix_data}, we find that productivity spillovers are comparable in strength (little, if any, shade gradient) across most Southeastern and coastal provinces and thus extend beyond the Shanghai and Guangzhou areas where the majority of the electric machinery manufacturing industry is concentrated. Therefore, the evidence of productivity spillovers that we find is not just a ``mechanical'' function of the spatial density of data. 

\paragraph{Endogenous Exposure to FDI.} The structural identification of our model only requires that the \textit{lagged} foreign equity share $G_{i,t-1}$ be weakly exogenous with respect to the future productivity innovation $\zeta_{it}$. Therefore, firms in our framework may experience endogenous updates to their exposure to foreign knowledge $G_{it}$ and even relocate based on the contemporaneous productivity $\omega_{it}$. Nonetheless, however mild an assumption, predeterminedness of $G_{i,t-1}$ may still be violated in case the firms\textemdash or their foreign investors\textemdash can forecast their future productivity (shocks). As a robustness check to this potential violation of the weak exogeneity of lagged foreign equity share, we re-estimate our baseline model using the inverse probability weighting (IPW) procedure and instrumentation. In our case, we need to weight only the second-stage regression since the first-stage share equation contains neither the FDI variable nor a productivity innovation. Analogously, instrumenting the lagged foreign equity share affects only the second stage. The results are summarized in Table \ref{tab:sp_dl_til__s_endo}.

% ---------------------------------------
\begin{table}[t]
	\centering
	\caption{Estimates of the Productivity Effects: Endogeneity of Lagged FDI}
	\label{tab:sp_dl_til__s_endo}
	\small
	\makebox[\linewidth]{
		\begin{tabular}{l c | ccc}
			\toprule
			& Exogenous	& IPW & IV-External  & IV-Lewbel  \\   
			\midrule
			\multicolumn{5}{c}{\textbf{---Median Estimates---}} \\[2pt]
			$SP$	& 0.327 & 0.477 & 0.466 & 0.414 \\
			& (0.211, 0.399) & (0.413, 0.512) & (0.411, 0.585) & (0.276, 0.515) \\
			$DL$	& 0.138 & 0.128 & 0.171 & 0.191 \\
			& (0.118, 0.164) & (0.105, 0.158) & (0.138, 0.241) & (0.053, 0.341) \\
			$TIL$	& 0.037 & 0.055 & 0.067 & 0.066 \\
			& (0.022, 0.050) & (0.044, 0.069) & (0.056, 0.100) & (0.019, 0.126) \\
			\midrule
			\multicolumn{5}{c}{\textbf{---Statistically $>0$ (\% Obs.)---}} \\[2pt]
			$SP$	& 83.84 & 98.55 & 97.17 & 79.56 \\
			$DL$	& 89.13 & 91.40 & 78.48 & 82.30 \\
			$TIL$	& 86.71 & 98.99 & 97.64 & 83.48 \\
			\midrule
			\multicolumn{5}{p{11cm}}{\footnotesize {\sc Notes:} Reported are the results for the productivity process in \eqref{eq:proddist_a_exp} under baseline specification. The two-sided 95\% bootstrap percentile confidence intervals for the median point estimates are in parentheses. Statistical positiveness is at the 5\% significance level, using one-sided bootstrap confidence intervals. The ``Exogenous'' column corresponds to our proposed estimation procedure under the structural assumption of weak exogeneity of $G_{i,t-1}$,  with the second-stage eq.~\eqref{eq:sst} estimated via least squares. The three models on the right allow for the violation of $\mathbb{E}[\zeta_{it}+\eta_{it}|G_{i,t-1}]=0$ and address the potential endogeneity of lagged FDI in \eqref{eq:sst} via (i) inverse probability weighting [IPW], (ii) instrumentation using external IVs including coastal province dummy, province-level openness measure and their interactions with predetermined $L_{i,t-1}$ [IV-External], (iii) instrumentation with the heteroskedasticity-based internal IV \`a la \citet{lewbel2012} [IV-Lewbel]. } \\
			\bottomrule[1pt]
		\end{tabular}
	}
\end{table}
% ---------------------------------------

By means of IPW, we seek to account for the potential selection (on observables) of firms by their foreign investors based on their \textit{future} productivity. We use the stabilized IPWs which are typically more numerically stable and produce narrow confidence bounds \citep{hernanrobins2019}. Also, note that we deal with a continuous ``treatment'' $G_{it}$ which is why our approach is different from the more conventional propensity score estimation suitable for binary treatments \citep[e.g.,][]{imbenswooldrdige2009}. The IPWs for a continuous treatment $G_{it}$ are given by $f_G(G_{it})/f_{G|\textbf{d}}(G_{it}|\textbf{d}_{i,t+1})$, where $f(\cdot)$ is a \textit{pdf}. [For more, also see \citet{hiranoimbens2004} and \citet{hernanrobins2019}.]
Note that  the vector of observables $\textbf{d}_{i,t+1}$ includes firm characteristics reflective of its performance next period, because of concern is the selection into treatment based on the future productivity.  We include the following correlates of the firm productivity that may influence its foreign exposure: size proxied by the logged labor, age, state equity share, government subsidy receipts, export intensity, normalized profits, the return on assets, leverage, logged total assets as well as the East coast dummy and the time trend.

To avoid the curse of dimensionality as well as the problem of near-zero extreme values associated with nonparametric estimation of densities, we employ a parametric maximum-likelihood approach to estimate $f_G$ and $f_{G|\textbf{d}}$. Given the bounded nature of a fractional variable $G_{it}\in[0,1]$, we assume it is Beta-distributed. Because beta distribution is not trivial to estimate, we impose few data-motivated restrictions on it. More specifically, we let $G_{it} \sim \text{Beta}(\alpha,\beta)$ and $G_{it}|\mathbf{d}_{i,t+1} \sim \text{Beta}(\alpha,\beta(\mathbf{d}_{i,t+1}))$ where, to match the data, we restrict the first shape parameter to a unit value ($\alpha=1$) and the second parameter/function $\beta(\cdot)$ to be greater than 1 so that the distribution of $G_{it}$ is unimodal with a zero mode in both instances.\footnote{Recall that the mode of the $\text{Beta}(\alpha,\beta)$ distribution is 0 for $0<\alpha\le 1$ and $\beta>1$.} The densities are estimated via maximum likelihood (ML), although the method of moments provides an alternative route. 

Since we seek to address the potential endogeneity of lagged $G_{i,t-1}$ in the second-stage least squares regression, we also lag the estimated IPWs to have them match the time period of ``treatment.'' In other words, we weight each observation in the second stage by  $\widehat{f}_G(G_{i,t-1})/\widehat{f}_{G|\textbf{d}}(G_{i,t-1}|\textbf{d}_{it})$.\footnote{More concretely, the IPW function is
\begin{align*}
	\frac{f_G(G_{i,t-1})}{f_{G|\textbf{d}}(G_{i,t-1}|\textbf{d}_{it})} = \frac{(1-G_{i,t-1})^{\beta_0-1}\Gamma(1+\beta_0)}{\Gamma(\beta_0)}\times\left[\frac{(1-G_{i,t-1})^{\beta(\textbf{d}_{it})-1}\Gamma(1+\beta(\textbf{d}_{it}))}{\Gamma(\beta(\textbf{d}_{it}))}\right]^{-1},
\end{align*}
where $\beta_0$ is a scalar shape parameter estimated via ML using $G_{it} \sim \text{Beta}(1,\beta_0)$; $\beta(\textbf{d}_{it})$ is a scalar function estimated via ML using $G_{it}|\textbf{d}_{i,t+1} \sim \text{Beta}(1,\beta(\textbf{d}_{i,t+1}))$ with $\beta(\textbf{d}_{it})$ parameterized using the exponential function of a single index; and $\Gamma(\cdot)$ is the Gamma function.} The results are summarized in the IPW column of Table \ref{tab:sp_dl_til__s_endo}.

Table \ref{tab:sp_dl_til__s_endo} also reports the estimates of productivity effects from the second stage estimated via generalized method of moments using external instruments (the IV-External column) and \citeauthor{lewbel2012}'s (2012) heteroskedasticity-based internal instruments (the IV-Lewbel column) for $G_{i,t-1}$. The external instruments include the East coast dummy, a province-level measure of openness (defined as the ratio of the sum of imports and exports to the gross domestic product) and their interactions with the firm-level lagged labor input. The two external instruments are motivated by the previous studies, such as \citet{eichengreenTong2007} and \citet{kelleryeaple2009}, and are selected to proxy friendly regional policies towards foreign capital, shipping costs and the overall ease of engaging in international trade and finance. We interact these instruments with the predetermined labor at the firm level to gain variation. For identification of the firm's productivity process based on heteroskedasticity, adapting \citet{lewbel2012} we first estimate an auxiliary equation for the endogenous regressor by regressing $G_{i,t-1}$ on all the other exogenous variables in the $\omega_{it}$ process, namely, $\omega_{i,t-1}$ and $\sum_{j \neq i} s_{ij,t-1}\omega_{j,t-1}$. The residuals from this auxiliary regression are then interacted with the demeaned $\omega_{i,t-1}$ and used to instrument for $G_{i,t-1}$. Here, we use the firm's predetermined lagged productivity $\omega_{i,t-1}$ as a ``$Z$'' variable that is uncorrelated with the \textit{product} of productivity innovation $\zeta_{it}$ (with which $G_{i,t-1}$ is suspected to be correlated) and the error in the auxiliary equation for $G_{i,t-1}$. For more details on instrumentation via heteroskedasticity, see \citet{lewbel2012}.

Controlling for potential endogeneity of the FDI exposure, we continue to find strong empirical evidence in support of significantly positive productivity spillovers for 80\% of firms or more. At least 78\% of manufacturers benefit from significant productivity boosts associated with receiving FDI. The changes in the effect sizes are not out of the ordinary either, and the rank correlation coefficient of the point estimate of productivity effects across these estimators is at least as high as 0.71. All in all, our findings remain qualitatively unchanged.

\paragraph{Bidimensional Spillovers.} In our main analysis, per the productivity process \eqref{eq:proddist_a_exp}, spillovers from all spatially proximate peers in the industry have the potential to affect the recipient firm's productivity in the same manner, no matter the exposure of these peers to foreign knowledge. Given the documented impact of FDI on firm productivity, it may also be of interest to allow for heterogeneity in the external productivity spillovers from the peers \textit{conditional} on their FDI status. To this end, we adapt our methodology to allow a more general evolution process for productivity that permits bidimensional spillovers by means of two spatiotemporal lags. Thus, we now consider the following evolution process of firm productivity:
\begin{equation}\label{eq:proddist_a_exp_2s}
	\omega_{it} = \mathbb{E}\left[\omega_{it} \Bigg|\ \omega_{i,t-1}, G_{i,t-1}, \sum_{j(\ne i)}s_{ij,t-1}^0\omega_{j,t-1}, \sum_{j(\ne i)}s_{ij,t-1}^1\omega_{j,t-1}\right] + \zeta_{it},
\end{equation}	
with the distinction between peer weights $\{s_{ijt}^0\}$ and  $\{s_{ijt}^1\}$ based on the peers' FDI status:
\begin{align}\label{eq:weight2}
	s_{ijt}^0= \frac{\mathbbm{1}\big\{G_{jt}=0 \ \text{and}\ (j,t)\in \mathcal{L}(i,t)\big\} }
	{\sum_{k(\ne i)=1}^{n} \mathbbm{1}\big\{G_{kt}=0 \ \text{and}\ (k,t)\in \mathcal{L}(i,t)\big\} }, \\
	s_{ijt}^1= \frac{\mathbbm{1}\big\{G_{jt}>0 \ \text{and}\ (j,t)\in \mathcal{L}(i,t)\big\} }
	{\sum_{k(\ne i)=1}^{n} \mathbbm{1}\big\{G_{kt}>0 \ \text{and}\ (k,t)\in \mathcal{L}(i,t)\big\} }.
\end{align}	

The cross-firm productivity spillovers are now bidimensional. The null intersection of the two sets of peers ensures separable identifiability of heterogeneous spillovers from (i) the wholly domestically owned peers $SP_{it}^0 = \partial \mathbb{E}[\omega_{it}|\cdot]\big/\partial \sum_{j(\ne i)}s_{ij,t-1}^0\omega_{j,t-1}$ and  (ii) foreign-invested peers $SP_{it}^1 = \partial \mathbb{E}[\omega_{it}|\cdot]\big/\partial \sum_{j(\ne i)}s_{ij,t-1}^1\omega_{j,t-1}$.

Table \ref{tab:sp_dl_til__ss} summarizes point estimates of bidimensional productivity spillovers under our baseline specification, whereby the firm's peer group $\mathcal{L}(i,t)$ is defined at the level of the same province and the entire 2-digit industry. Just like in the case of our main model with the unidimensional cross-firm dependence, we continue to find substantial and positive productivity spillovers in the industry. However, by disentangling the peer effects of wholly domestically owned and foreign-invested neighbors, we find that the former group has a significantly larger effect on its neighbors. The median spillover elasticity from fully domestic firms $SP^0$ is 0.30, whereas the counterpart estimate $SP^1$ from the foreign-invested peers is 0.17 only. Regardless of the peer group, the productivity spillovers are statistically positive for the overwhelming majority of firms in the industry (83\% or more). This is on par with the extent of spillovers that we have found in our main model with unidimensional spillovers. 

% ---------------------------------------
\begin{table}[t]
	\centering
	\caption{Estimates of Bidimensional Productivity Spillovers}
	\label{tab:sp_dl_til__ss}
	\small
	\makebox[\linewidth]{
		\begin{tabular}{l ccc|c}
			\toprule
			& \multicolumn{3}{c}{\textit{Point Estimates}} & {\textit{Statistically $>0$}} \\   
			Estimand & 1st Qu. & Median & 3rd Qu. & (\% Obs.)  \\
			\midrule
			$SP^0$ 	& 0.153 & 0.300 & 0.428 & 82.58 \\
			& (0.090, 0.180) & (0.198, 0.349) & (0.302, 0.504) &  \\[2pt]
			$SP^1$ 	& 0.158 & 0.172 & 0.185 & 98.89 \\
			& (0.119, 0.188) & (0.129, 0.204) & (0.137, 0.216) &  \\
			\midrule
			\multicolumn{5}{p{12cm}}{\footnotesize {\sc Notes:} Reported are the semiparametric estimates of bidimensional productivity spillovers from \eqref{eq:proddist_a_exp_2s} under our baseline specification. The left panel summarizes point estimates of $SP_{it}^0$ and $SP_{it}^1$ with the corresponding two-sided 95\% bootstrap percentile confidence intervals in parentheses. The last column reports the share of observations for which the point estimates are statistically positive at the 5\% significance level using one-sided bootstrap percentile confidence intervals.} \\
			\bottomrule[1pt]
		\end{tabular}
	}
\end{table}
% ---------------------------------------

The documented heterogeneity in the magnitudes of spillovers across the two types of peers suggests that it is relatively ``easier'' for Chinese manufacturers to learn from other domestic firms that are \textit{not} recipients of FDI. This may be because foreign-invested firms are more protective of their newly adopted foreign technologies/knowledge, which makes it more difficult to learn from them. At the same time, it also may be that learning from these foreign-invested firms is simply more difficult because their practices are too advanced and biased towards more productive/efficient firms in the first place. If so, firms that are already foreign-invested\textemdash and, hence, are more productivity due to direct learning effects of FDI\textemdash are to enjoy larger spillovers from their peers who have also received foreign investments. This is corroborated by the evidence in Table \ref{tab:sp_dl_til__ss_hetero}: the coefficient on the firms' \textit{own} FDI exposure is negative for $SP^0$ and positive for $SP^1$.

% -----------------------------------
\begin{table}[t]
	\caption{Heterogeneity in Bidimensional Productivity Spillovers}
	\label{tab:sp_dl_til__ss_hetero}
	\centering	\small
	\begin{tabular}{lcc}
		\toprule[1pt]
		& $SP^0$ & $SP^1$  \\
		\midrule
		$\omega_{i,t-1} $ 			& --0.720 & --0.070 \\
		& (--0.882, --0.571) & (--0.125, --0.019) \\
		$G_{i,t-1}$ 						& --0.185 & 0.036 \\
		& (--0.355, --0.073) & (--0.151, 0.130) \\
		$\sum_j s_{ij,t-1}^0\omega_{j,t-1}$ & 1.206 & --0.147 \\
		& (0.591, 1.459) & (--0.276, --0.023) \\
		$\sum_j s_{ij,t-1}^1\omega_{j,t-1}$ & --0.147 & 0.144 \\
		& (--0.276, --0.023) & (0.101, 0.178) \\
		\midrule
		\multicolumn{3}{p{8.4cm}}{\footnotesize {\sc Notes:} Reported are the parameter estimates for the $SP^0$ and $SP^1$ functions derived from the polynomial approximation of the conditional mean of $\omega_{it}$ in the productivity process formulation with bidimensional spillovers in \eqref{eq:proddist_a_exp_2s}. Two-sided 95\% bootstrap percentile confidence intervals in parentheses. These correspond to our baseline specification, with (i) each firm's peers restricted to the firms located in the same province and the industrial scope of spillovers defined at the level of the entire 2-digit industry, (ii) the technical change flexibly controlled for using a series of year effects. }\\
		\bottomrule[1pt]
	\end{tabular}
	
\end{table} 
% -----------------------------------

The results in Table \ref{tab:sp_dl_til__ss_hetero} continue to indicate that the more productivity firms have less absorptive capacity to learn from their peers. For both the $SP^0$  and $SP^1$ spillovers, the effect size increases with the average productivity of peers from whom spillovers originate. But on the other hand, the strength of spillovers from one peer group declines with the average productivity of the other group (note the negative coefficient on the cross-peer productivity). Taken together, these two findings suggest some substitutability between learning from the two groups along with the recipient firm's finite capacity to absorb such spillovers from the peers in a given period. Thus, if the foreign-invested peers improve their productivity, the firm starts learning more from them and less from the non-foreign-invested peers, and vice versa.

To conclude, although we find evidence of heterogeneity in the strength of spillovers from wholly-domestic versus foreign-invested peers (with those from the former being relatively stronger), in the grand scheme of things, our main findings stay the same: productivity spillovers are positive and significant for most firms in the industry. 

\paragraph{Asymmetric Spillovers.} As we explain in Appendix \ref{sec:appendix_asym}, the peer weighing scheme that we use in our analysis treats cross-firm spillovers symmetrically in that all members of a peer group affect each other's productivity. That is, each $i$th firm's productivity is influenced by the average productivity of all its peers: those that are more \textit{and} those that are less productive that the firm $i$ itself. But should one choose to regulate the direction of productivity spillovers by restricting them to occur \textit{from} more productive \textit{to} less productive firms, our framework can be modified to accommodate that, albeit with additional timing assumptions. 

% ---------------------------------------
\begin{table}[t]
	\centering
	\caption{Estimates of the Productivity Effects with Asymmetric Spillovers}
	\label{tab:asym}
	\footnotesize
	\makebox[\linewidth]{
		\begin{tabular}{l ccc|c}
			\toprule
			& \multicolumn{3}{c}{\textit{Point Estimates}} & {\textit{Statistically $>0$}} \\   
			Estimand & 1st Qu. & Median & 3rd Qu. & (\% Obs.)  \\
			\midrule
			$SP$ 	& 0.301 & 0.331 & 0.362 & 99.54 \\
			& (0.277, 0.328) & (0.306, 0.361) & (0.335, 0.400) &  \\
			$DL$ 	& 0.083 & 0.133 & 0.169 & 87.01 \\
			& (0.058, 0.106) & (0.104, 0.156) & (0.135, 0.192) &  \\	
			\midrule
			\multicolumn{5}{p{11.1cm}}{\footnotesize {\sc Notes:} Reported are the results based on the productivity process formulation with asymmetric productivity spillovers in \eqref{eq:proddist_a_exp*} using our baseline specification of $\mathcal{L}(i,t)$. The left panel summarizes point estimates of $SP_{it}$ and $DL_{it}$ with the corresponding two-sided 95\% bootstrap percentile confidence intervals in parentheses. The last column reports the share of observations for which the point estimates are statistically positive at the 5\% significance level using one-sided bootstrap percentile confidence intervals.} \\
			\bottomrule[1pt]
		\end{tabular}
	}
\end{table}
% ---------------------------------------

We estimate such an asymmetric specification given in \eqref{eq:proddist_a_exp*}. Table \ref{tab:asym} summarizes the corresponding estimates of productivity effects. Comparing these results with our main estimates in Table \ref{tab:sp_dl_til__s}, we see that the estimates of direct learning stay by and large unchanged, as expected. While comparable at the median, the asymmetric spillover effect estimates exhibit a much smaller variation in effect size (perhaps, because the peer pool is more homogeneous now) but are \textit{as} prevalent as they are when we model them symmetrically.

% ------------------------------------------------------------------------------------------
% ------------------------------------------------------------------------------------------

\section*{References}

{\onehalfspacing \small \parskip 5pt

\noindent Ackerberg, D. A., Caves, K., \& Frazer, G. (2015). Identification properties of recent production function estimators. \textit{Econometrica}, 83, 2411--2451.

\noindent Bramoull\'e, Y., Djebbari, H., \& Fortin, B. (2009). Identification of peer effects through social networks. \textit{Journal of Econometrics}, 150, 41--55.

\noindent De Loecker, J. (2013). Detecting learning by exporting. \textit{American Economic Journal: Microeconomics}, 5, 1--21.

\noindent De Loecker, J., Goldberg, P. K., Khandelwal, A. K., \& Pavcnik, N. (2016). Prices, markups, and trade reform. \textit{Econometrica}, 84, 445--510.

\noindent De Loecker, J. \& Warzynski, F. (2012). Markups and firm-level export status. \textit{American Economic Review}, 102, 2437--2471.

\noindent Deloitte China Manufacturing Industry Group (2013). A new stage for overseas expansion for China's equipment manufacturing industry. Report by Deloitte China Research and Insight Centre.

\noindent Doraszelski, U. \& Jaumandreu, J. (2013). R\&D and productivity: Estimating endogenous productivity. \textit{Review of Economic Studies}, 80, 1338--1383.

\noindent Doraszelski, U. \& Jaumandreu, J. (2018). Measuring the bias of technological change. \textit{Journal of Political Economy}, 126, 1027--1084.

\noindent Euro Exim Bank (2020). Export of electrical machinery from China. Euro Exim Bank Global Finance Blog; October 30, 2020.

\noindent Efron, B. (1987). Better bootstrap confidence interval. \textit{Journal of American Statistical Association}, 82, 171--200.

\noindent Eichengreen, B. \& Tong, H. (2007). Is China's FDI coming at the expense of other countries? \textit{Journal of the Japanese and International Economies}, 21(2), 153--172.

\noindent Gandhi, A., Navarro, S., \& Rivers, D. (2020). On the identification of gross output production functions. \textit{Journal of Political Economy}, 128, 2973--3016.

\noindent Griliches, Z. (1979). Issues in assessing the contribution of research and development to productivity growth. \textit{Bell Journal of Economics}, 10, 92--116.

\noindent Griliches, Z. \& Mairesse, J. (1998). Production functions: The search for identification. In \textit{Econometrics and Economic Theory in the Twentieth Century: The Ragnar Frisch Centennial Symposium} (pp. 169--203). Cambridge University Press.

\noindent Hahn, J., Liao, Z., \& Ridder, G. (2018). Nonparametric two-step sieve M estimation and inference. \textit{Econometric Theory}. forthcoming.

\noindent Hern\'an, M. A. \& Robins, J.M. (2020). \textit{Causal Inference: What If.} Boca Raton: Chapman \& Hall.

\noindent Hirano, K. \& Imbens, G.W. (2004). The propensity score with continuous treatments. In W. A. Shewhart \& S. S.Wilks (Eds.), \textit{Applied Bayesian Modeling and Causal Inference from Incomplete-Data Perspectives: An Essential Journey with Donald Rubin's Statistical Family.} New York: Wiley \& Sons, Ltd.

\noindent Horowitz, J. L. (2001). The bootstrap. In J. J. Heckman \& E. Leamer (Eds.), \textit{Handbook of Econometrics} (5 ed.). chapter 52, (pp. 3159--3228). Elsevier Science B.V.

\noindent Ihrcke, J. \& Becker, K. (2006). Study on the future opportunities and challenges of EU-China trade and investment relations. Study 1 of 12: Machinery. Report commissioned and financed by the Commission of the European Communities.

\noindent Imbens, G.W. \& Wooldridge, J. M. (2009). Recent developments in the econometrics of program evaluation. \textit{Journal of Economic Literature}, 47, 5--86.

\noindent Kelejian, H. H. \& Prucha, I. R. (1999). A generalized moment estimator for the autoregressive parameter in a spatial model. \textit{International Economic Review}, 40, 509--533.

\noindent Keller, W. \& Yeaple, S. R. (2009). Multinational enterprises, international trade, and productivity growth: Firm-level evidence from the United States. \textit{Review of Economics and Statistics}, 91, 821--831.

\noindent Kuersteiner, G. M. \& Prucha, I. R. (2020). Dynamic spatial panel models: Networks, common shocks, and sequential exogeneity. \textit{Econometrica}, 88, 2109--2146.

\noindent Lee, L.-f. (2007). GMM and 2SLS estimation of mixed regressive, spatial autoregressive models. \textit{Journal of Econometrics}, 137, 489--514.

\noindent Lewbel, A. (2012). Using heteroskedasticity to identify and estimate mismeasured and endogenous regressor models. \textit{Journal of Business and Economic Statistics}, 30, 67--80.

\noindent Mammen, E. (1993). Bootstrap and wild bootstrap for high dimensional linear models. \textit{Annals of Statistics}, 21, 255--285.

\noindent Manski, C. F. (1993). Identification of endogenous social effects: The reflection problem. \textit{Review of Economic Studies}, 60, 531--542.

\noindent Newey, W. K. (1984). A method of moments interpretation of sequential estimators. \textit{Economics Letters}, 14(2), 201--206.

\noindent Shao, J. \& Tu, D. (1995). \textit{The Jackknife and Bootstrap.} Springer-Verlag New York Inc.

}

% ------------------------------------------------------------------------------------------
% ------------------------------------------------------------------------------------------
% ------------------------------------------------------------------------------------------
% ------------------------------------------------------------------------------------------
% ------------------------------------------------------------------------------------------

\end{document}